# WHAT ARE "X-SHAPED" RADIO SOURCES TELLING US? I. VERY LARGE ARRAY IMAGING OF A LARGE SAMPLE OF LOW AXIAL RATIO RADIO SOURCES

David H. Roberts

*Department of Physics MS-057, Brandeis University, Waltham, MA 02453-0911, USA*

Lakshmi Saripalli

*Raman Research Institute, C. V. Raman Avenue, Sadashivanagar, Bangalore 560080, India*

Kevin X. Wang

*Department of Physics MS-057, Brandeis University, Waltham, MA 02453-0911, USA*

Mayuri Sathyanarayana Rao & Ravi Subrahmanyan

*Raman Research Institute, C. V. Raman Avenue, Sadashivanagar, Bangalore 560080, India*

Carly C. KleinStern, Christopher Y. Morii-Sciolla, & Liana Simpson

*Department of Physics MS-057, Brandeis University, Waltham, MA 02453-0911, USA*

roberts@brandeis.edu, lsarapal@rri.res.in

## ABSTRACT

We present historical VLA and Jansky Very Large Array multi-frequency multi-array radio continuum imaging of a unique sample of 100 radio sources that have been selected on the basis of low axial ratios. These observations allow us the opportunity to study radio sources with synchrotron plasma that is significantly offset from the main radio axis and therefore open a window into investigations of physical mechanisms responsible for depositing the plasma in off-axis regions. These images are discussed in detail in subsequent papers in this series (Saripalli & Roberts 2017; Roberts & Saripalli 2017).

*Subject headings:* black holes — galaxies: active — radio continuum: galaxies

Version of August 7, 2017.

## 1. INTRODUCTION

The "X-shaped" radio galaxies (XRGs) are of intense interest because one possible scenario for their formation involves pairs of supermassive black holes (SMBHs). Such pairs are presumed to be the natural results of galaxy mergers (Begelman, Blandford, & Rees 1980). This led Merritt & Ekers (2002) to suggest that XRGs were signposts for SMBH pairs. Since the coalescence of SMBH pairs is thought to be the leading source of the low-frequency gravitational wave background, their statistics can give us a handle on the rate of such events. In Roberts et al. (2015), henceforth Paper I, we presented the beginning of our study of XRGs, archival VLA imaging of a subsample of 100 candidate XRGs selected by Cheung (2007). There we showed images of 52 of the candidates, and discussed the various morphologies that we found. We concluded in Roberts, Saripalli, & Subrahmanyan (2015) that the rate of formation of binary SMBHs may have been previously overestimated by a factor of 3-5.

In this paper we present Jansky Very Large Array multi-frequency multi-array observations of 86 radio galaxies and quasars out of the Cheung (2007) sample (the rest were missed due to insufficient observing time). Including the archival data presented by Roberts et al. (2015), we made images of 95/100 of the candidate XRGs. Subsequent papers in this series discuss the results in detail (Saripalli & Roberts 2017; Roberts & Saripalli 2017).

The next section gives details of the observations and imaging methods. Section 3 presents the results, followed by a summary in Section 4.

## 2. OBSERVATIONS

We observed the Cheung (2007) sample of sources from 2016 February 04 through 2017 January 10 using the NRAO Jansky Very Large Array at L- (1.4 GHz) and S-bands (3 GHz) and the A, B, and C-arrays (one epoch was in the transitional AB-array). Each source was observed in a single snapshot that ranged from about 180 to about 240 seconds duration, depending on scheduling constraints. At S band we used 16 spectral windows of 128 MHz with a net average frequency of about 3 GHz, while at L-band we used 16 spectral windows of 64 MHz width, with an average frequency of between 1.4 and 1.6 GHz, depending on the flagged spectral windows. The spectral windows were each divided into 64 channels. Each data set was edited for interference by the pipeline (see below) and then inspected by hand, with a typical loss of a few of the spectral windows. The log of our observations is given in Table 1. All-in-all we had data from both archival VLA data (Roberts et al. 2015) and Jansky VLA data on 46 sources, archival data only on six, Jansky VLA data only on 43, and a total of five sources were missed altogether.

Although JVLA observations were proposed for the full sample of 99 sources (one source was found to be of head-tail nature based on the archival data in Roberts et al. (2015)), we could observe only a subset of sources in each of the arrays because of the "C" priority assigned to most of the granted observing time. The data were analyzed at Brandeis University. We used our own version of the CASA (McMullin et al. 2007) data calibration pipeline package which we edited to add polarization calibration. Total intensity, spectral index, and complex polarization images were made of every source-frequency-array combination using multifrequency synthesis (Rao & Cornwell 2011) in CASA. Where possible the images were self-calibrated in CASA using a Python script to do the bookkeeping (Harrrison 2014).[1]

## 3. RESULTS

Radio images for all of the sources are presented in sets of nine in Figures 1–27. For each source the various images are presented in order of increasing angular resolution–S-band C-array, L/B, S/B, C/B, and L/A. The lowest contour level and peak for each image are presented in Table 2. In every plot the contour levels increase by successive factors of $2^{1/2}$. Where we had both archival and JVLA images at L-band in the A-array we present only the JVLA image; otherwise we present the archival image (see Table 1). Optical overlays using the Sloan Digital Sky Survey (DR12) r-band images are presented in Figures 28–38; for each source we used what we judged to be the best VLA image. In a few cases SDSS images were not available, and there we used the USNO Image and Catalog Archive Server (http://www.usno.navy.mil/USNO/astrometry/optical-IR-prod/icas) to find suitable optical images.

## 4. SUMMARY

In this paper we have presented Jansky Very Large Array multi-band multi-array continuum imaging of a unique sample of 100 radio galaxies that have been selected on the basis of low axial ratios. In our next paper we select a sample of 87 for study, thoroughly characterize their morphologies, and present scenario for their formation (Saripalli & Roberts 2017). In Roberts & Saripalli (2017) we discuss the sources that show clear evidence of black hole axis precession. **?** uses polarization and spectral index information to examine models

---

[1]The self-calibration script emulates the flow of the Caltech program DIFMAP; it is available from D.H.R.

for the formation of the structure of radio galaxies with significant off-axis emission.

In the future we will also be deriving radio-optical relationships for those sources for which suitable images of the host galaxies exist. We also intend to obtain more complete optical information on the host galaxies by carrying out comprehensive radio-optical studies for the sample.

## 5.   ACKNOWLEDGMENTS

The National Radio Astronomy Observatory is a facility of the National Science Foundation, operated under cooperative agreement by Associated Universities, Inc. Funding for the Sloan Digital Sky Survey IV has been provided by the Alfred P. Sloan Foundation, the U.S. Department of Energy Office of Science, and the Participating Institutions. SDSS-IV acknowledges support and resources from the Center for High-Performance Computing at the University of Utah. The SDSS web site is www.sdss.org. D. H. R. gratefully acknowledges the support of the William R. Kenan, Jr. Charitable Trust. We thank Rachel Harrison for her CASA self-calibration script.

*Facilities:* Historical VLA Data Archive (see Paper I); Jansky VLA (VLA 16A-220, 16B-023).

## 6. APPENDIX–ALL OF THE IMAGES

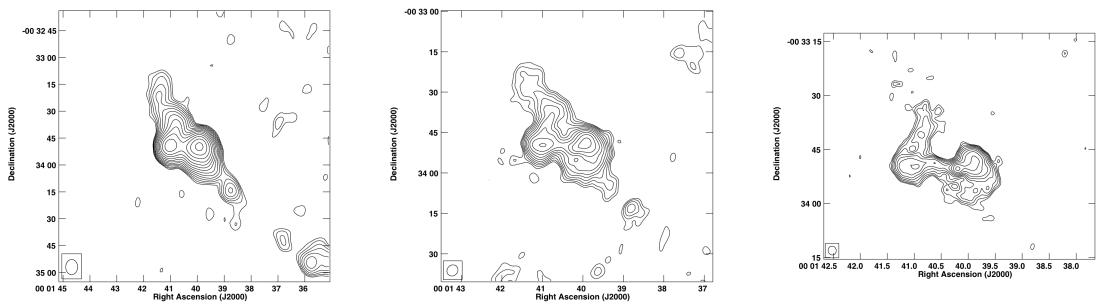

(a) J0001−0033, C-array S-band.(b) J0001−0033, B-array L-band.(c) J0001−0033, B-array S-band.

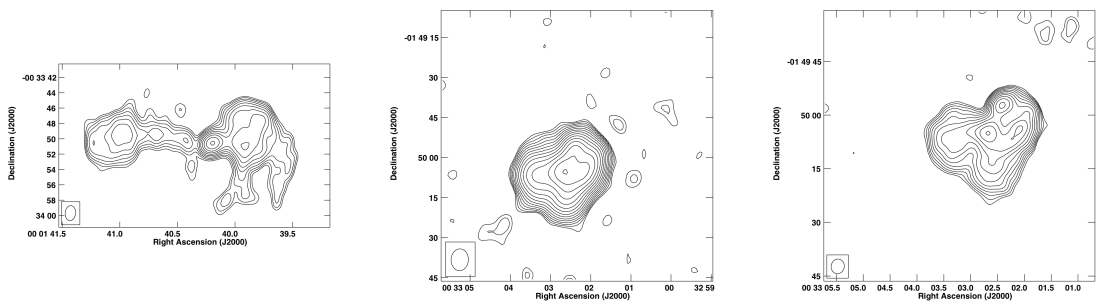

(d) J0001−0033, A-array L-band.(e) J0033−0149, C-array S-band.(f) J0033−0149, B-array L-band.

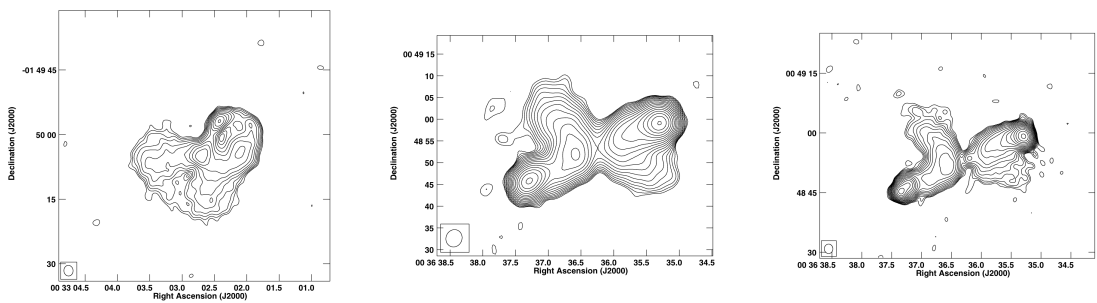

(g) J0033−0149, B-array S-band.(h) J0036+0048, B-array L-band.(i) J0036+0048, B-array S-band.

Fig. 1.—: VLA images of a sample of 100 low axial ratio radio sources (1/27).

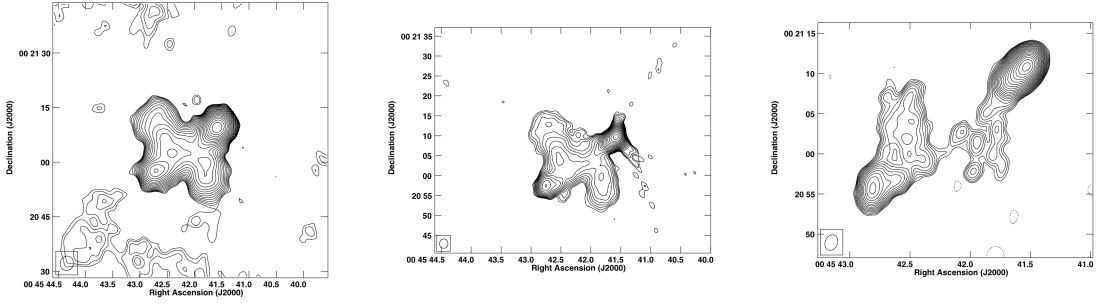

(a) J0045+0021, B-array L-band.(b) J0045+0021, B-array S-band.(c) J0045+0021, B-array C-band.

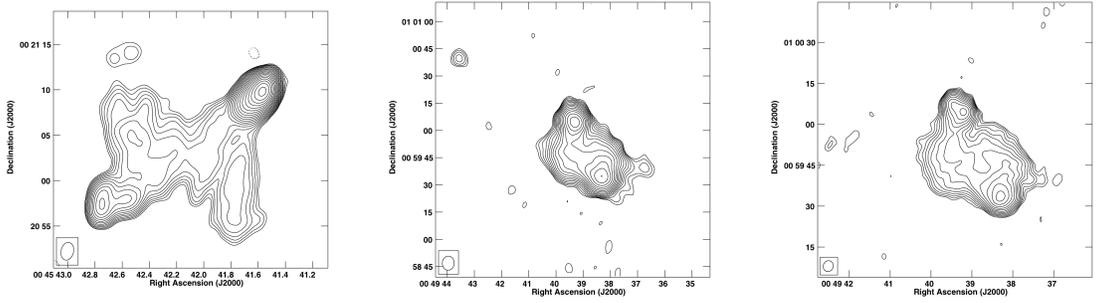

(d) J0045+0021, A-array L-band.(e) J0049+0059, C-array S-band.(f) J0049+0059, B-array L-band.

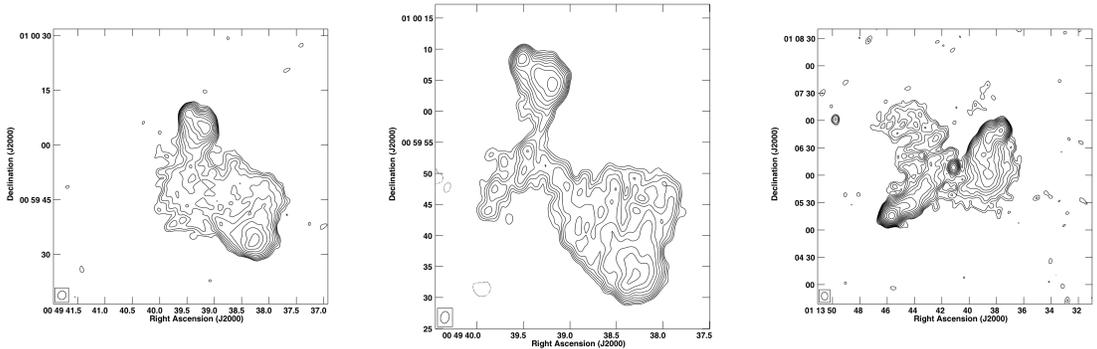

(g) J0049+0059, B-array S-band.(h) J0049+0059, A-array L-band.(i) J0113+0106, C-array S-band.

Fig. 2.—: VLA images of a sample of 100 low axial ratio radio sources (2/27).

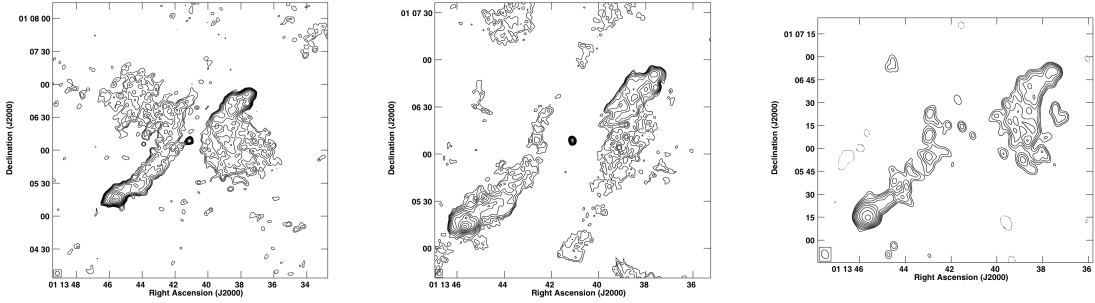

(a) J0113+0106, B-array L-band.(b) J0113+0106, B-array S-band.(c) J0113+0106, A-array L-band.

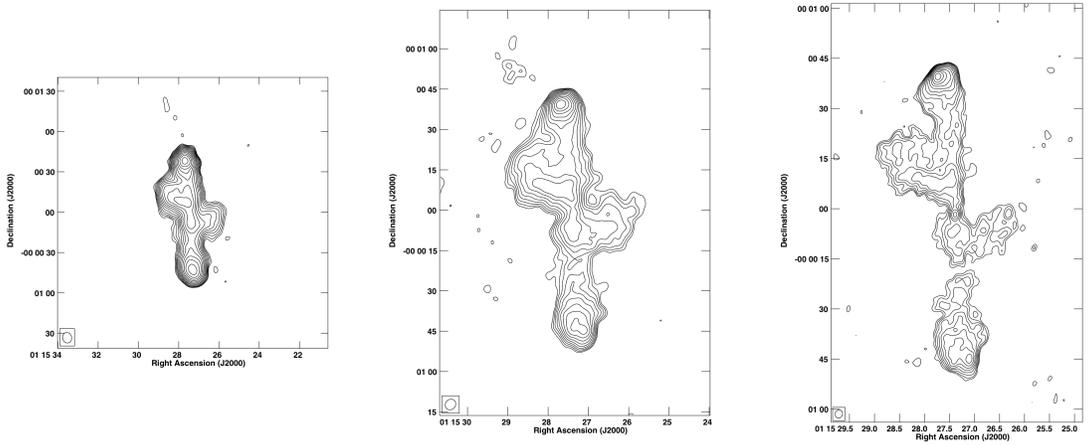

(d) J0115−0000, C-array S-band.(e) J0115−0000, B-array L-band.(f) J0115−0000, B-array S-band.

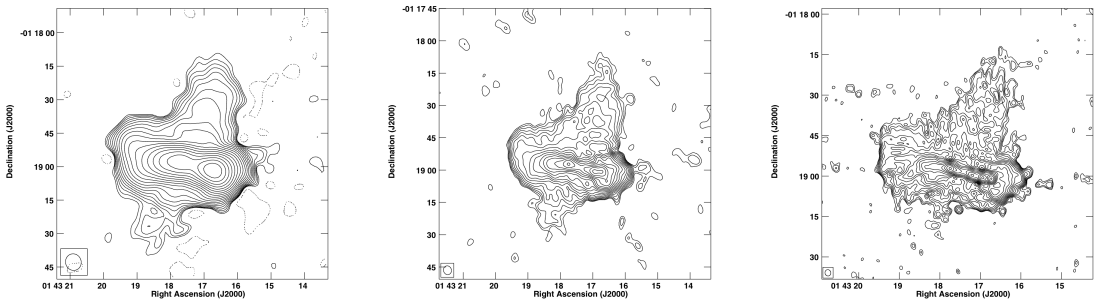

(g) J0143−0119, C-array S-band.(h) J0143−0119, B-array L-band.(i) J0143−0119, B-array S-band.

Fig. 3.—: VLA images of a sample of 100 low axial ratio radio sources (3/27).

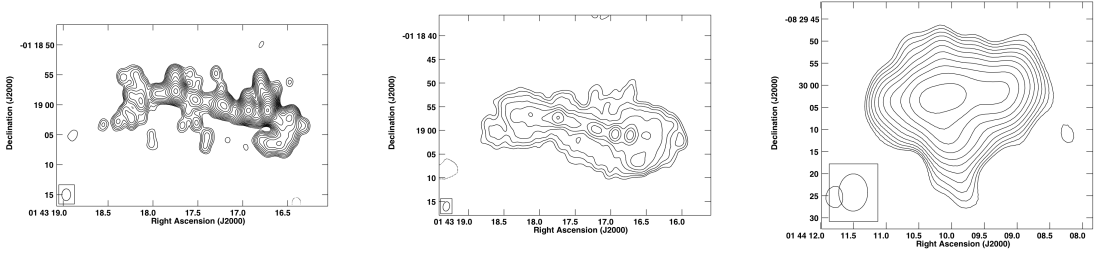

(a) J0143−0119, B-array C-band. (b) J0143−0119, A-array L-band. (c) J0144−0830, C-array S-band.

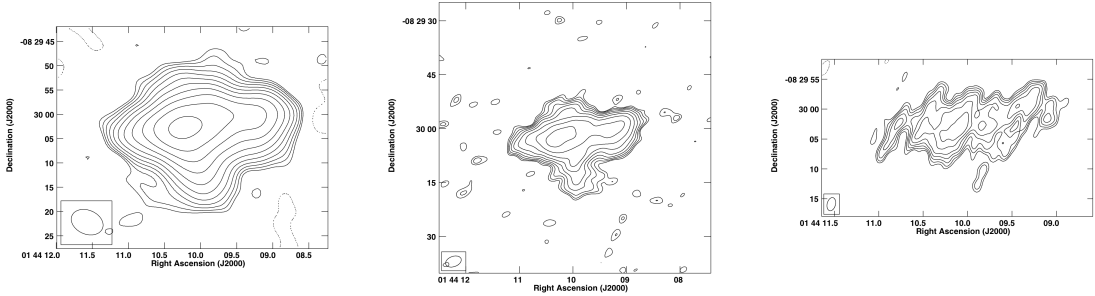

(d) J0144−0830, B-array L-band. (e) J0144−0830, B-array S-band. (f) J0144−0830, A-array L-band.

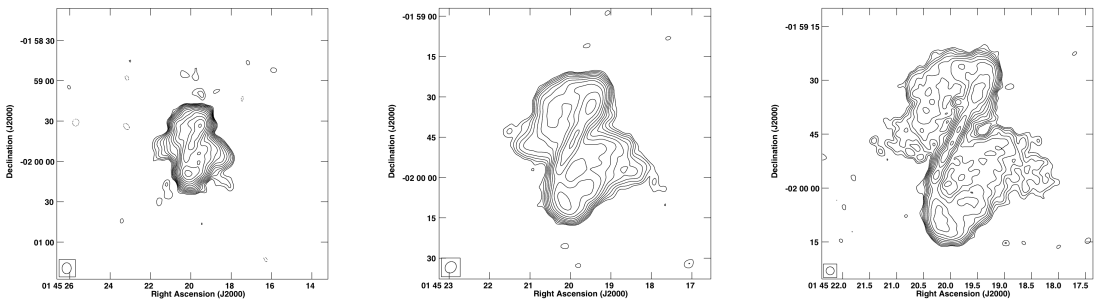

(g) J0145−0159, C-array S-band. (h) J0145−0159, B-array L-band. (i) J0145−0159, B-array S-band.

Fig. 4.—: VLA images of a sample of 100 low axial ratio radio sources (4/27).

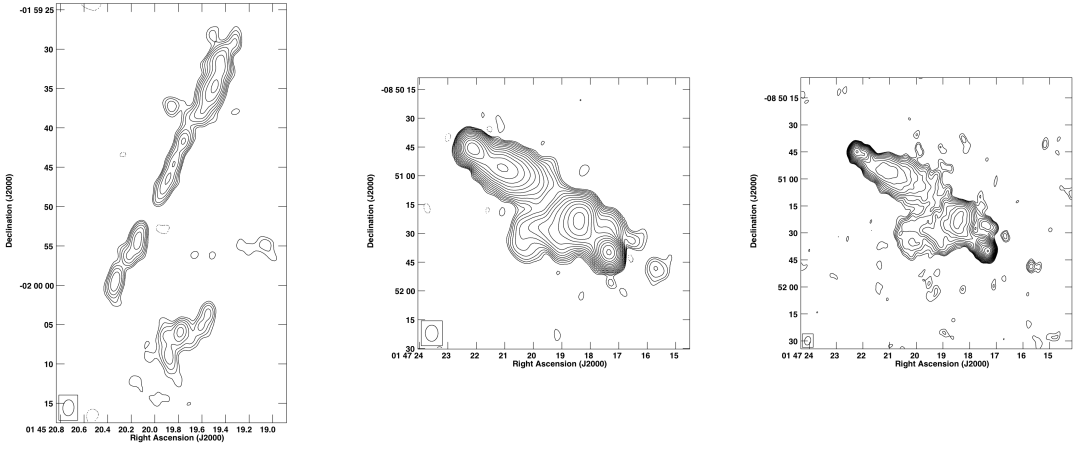

(a) J0145−0159, A-array L-band.(b) J0147−0851, C-array S-band.(c) J0147−0851, B-array L-band.

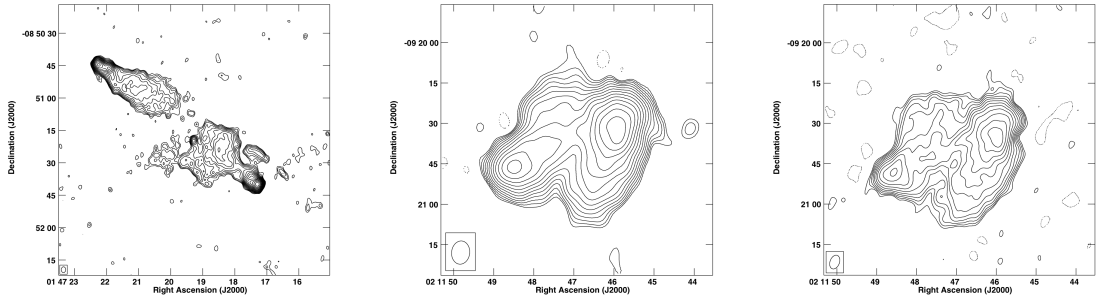

(d) J0147−0851, B-array S-band.(e) J0211−0920, C-array S-band.(f) J0211−0920, B-array L-band.

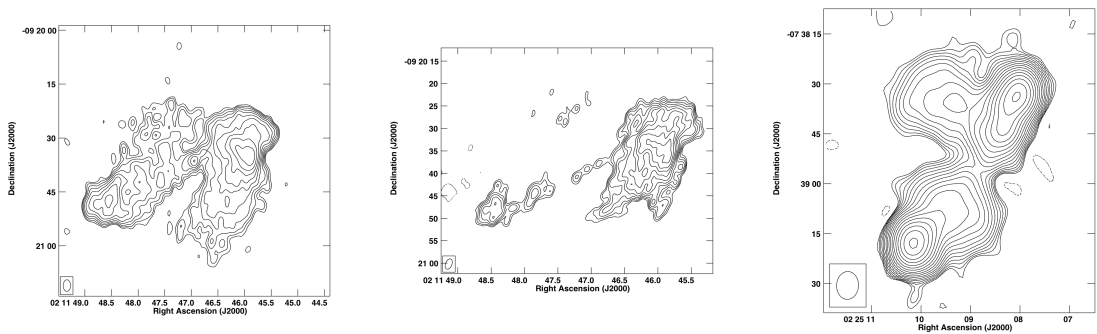

(g) J0211−0920, B-array S-band.(h) J0211−0920, A-array L-band.(i) J0225−0738, C-array S-band.

Fig. 5.—: VLA images of a sample of 100 low axial ratio radio sources (5/27).

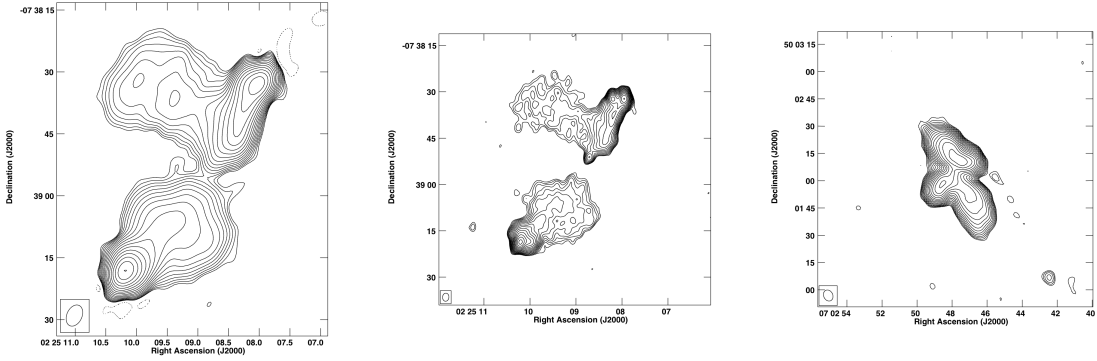

(a) J0225−0738, B-array L-band.(b) J0225−0738, B-array S-band.(c) J0702+5002, C-array S-band.

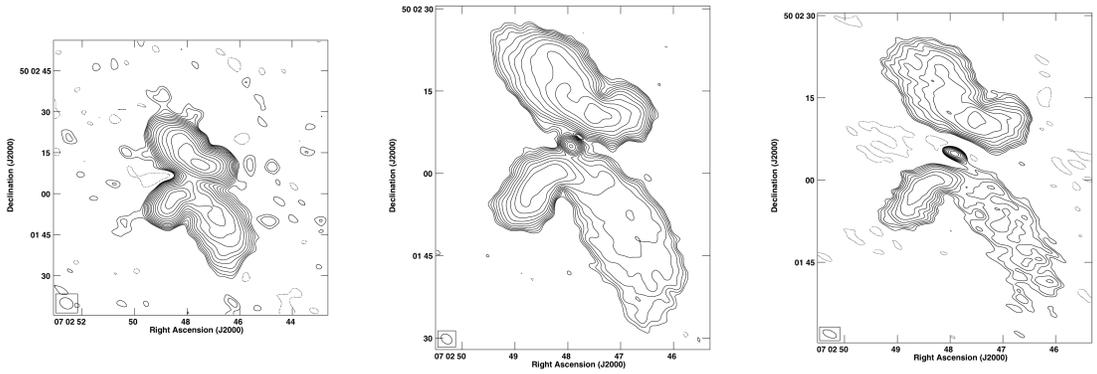

(d) J0702+5002, B-array L-band.(e) J0702+5002, B-array S-band.(f) J0702+5002, A-array L-band.

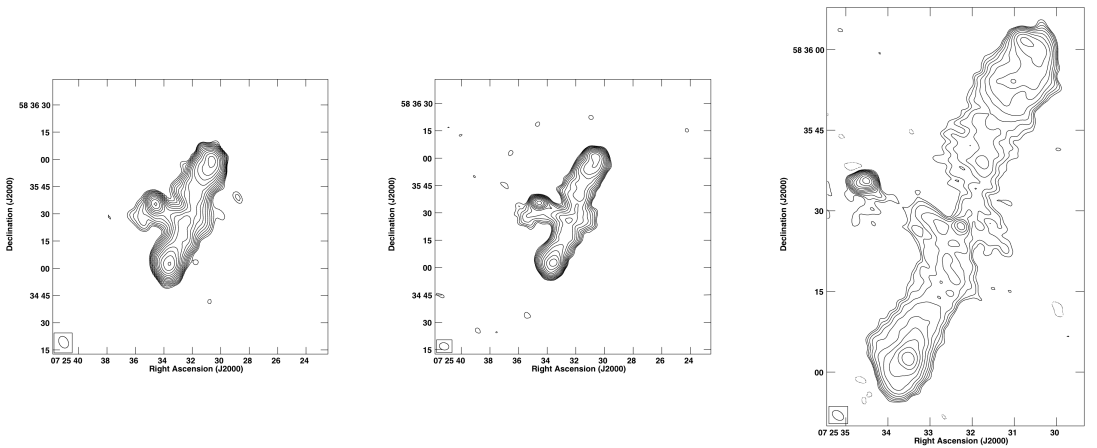

(g) J0725+5835, C-array S-band.(h) J0725+5835, B-array L-band.(i) J0725+5835, B-array S-band.

Fig. 6.—: VLA images of a sample of 100 low axial ratio radio sources (6/27).

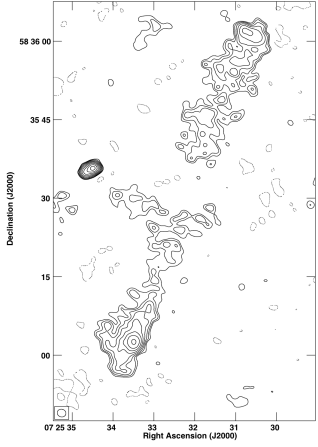 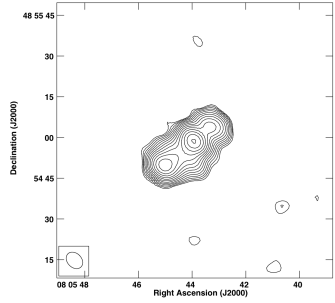 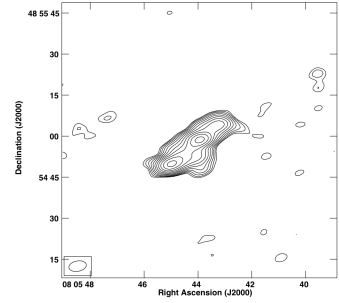

(a) J0725+5835, A-array L-band.(b) J0805+4854, C-array S-band.(c) J0805+4854, B-array L-band.

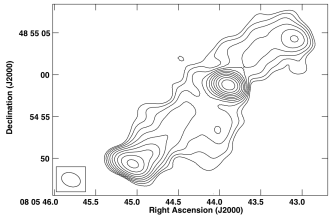 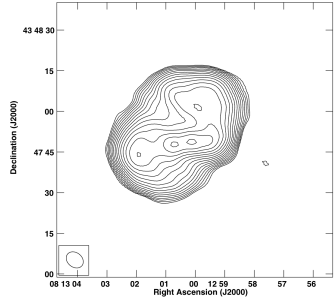 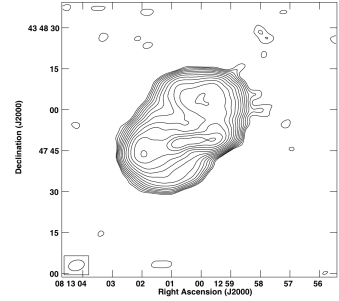

(d) J0805+4854, B-array S-band.(e) J0813+4347, C-array S-band.(f) J0813+4347, B-array L-band.

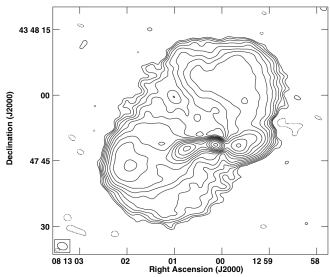 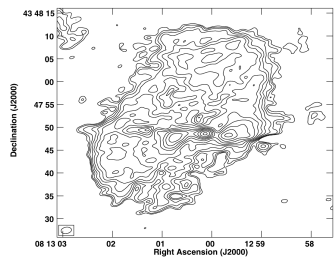 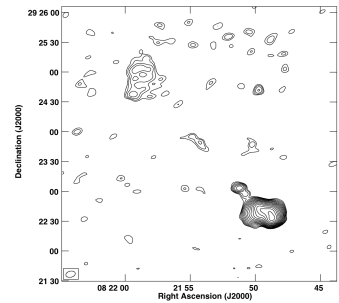

(g) J0813+4347, B-array S-band.(h) J0813+4347, A-array L-band.(i) J0821+2922, C-array S-band.

Fig. 7.—: VLA images of a sample of 100 low axial ratio radio sources (7/27).

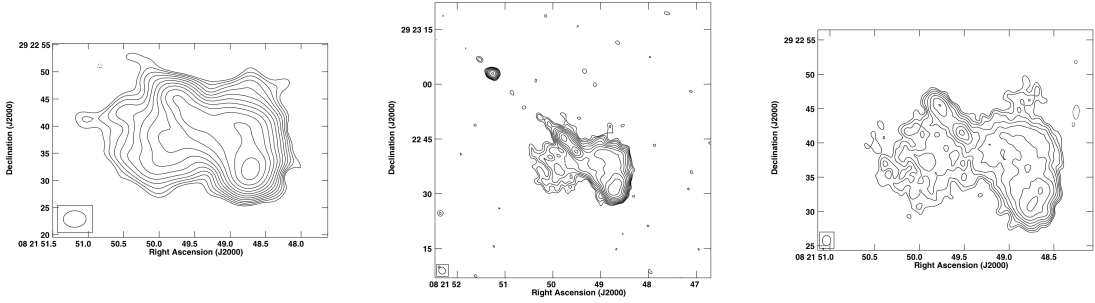

(a) J0821+2922, B-array L-band. (b) J0821+2922, B-array S-band. (c) J0821+2922, A-array L-band.

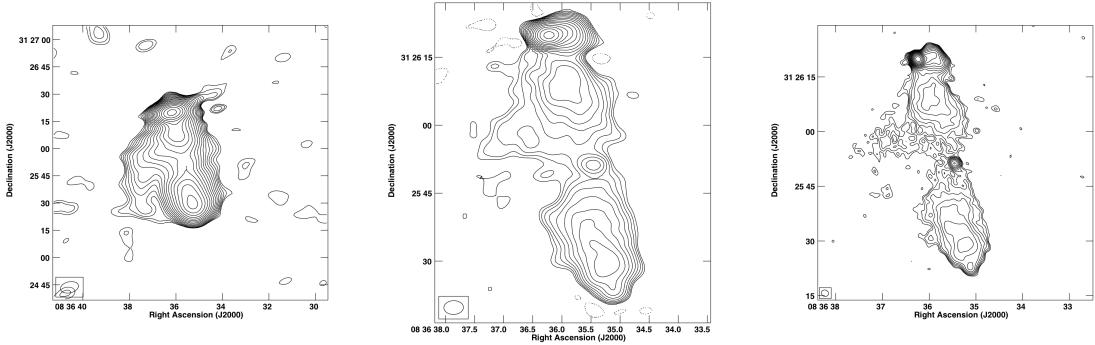

(d) J0836+3125, C-array S-band. (e) J0836+3125, B-array L-band. (f) J0836+3125, B-array S-band.

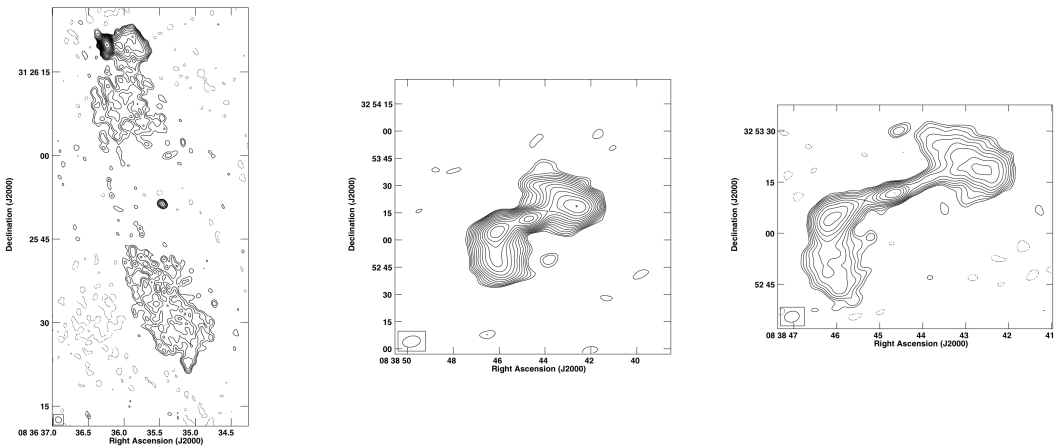

(g) J0836+3125, A-array L-band. (h) J0838+3253, C-array S-band. (i) J0838+3253, B-array L-band.

Fig. 8.—: VLA images of a sample of 100 low axial ratio radio sources (8/27).

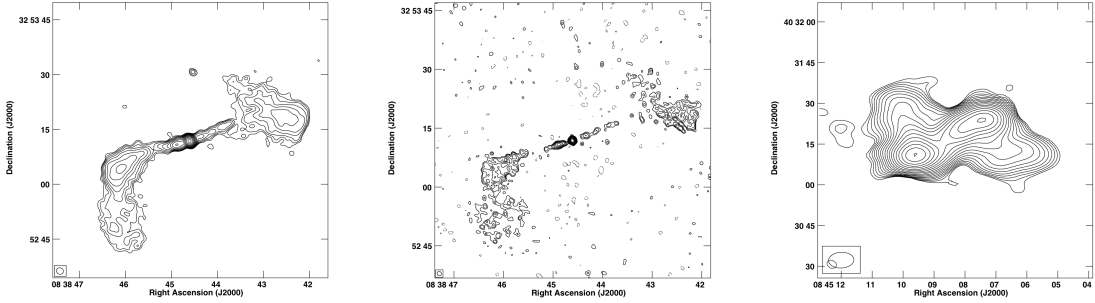

(a) J0838+3253, B-array S-band.(b) J0838+3253, A-array L-band.(c) J0845+4031, C-array S-band.

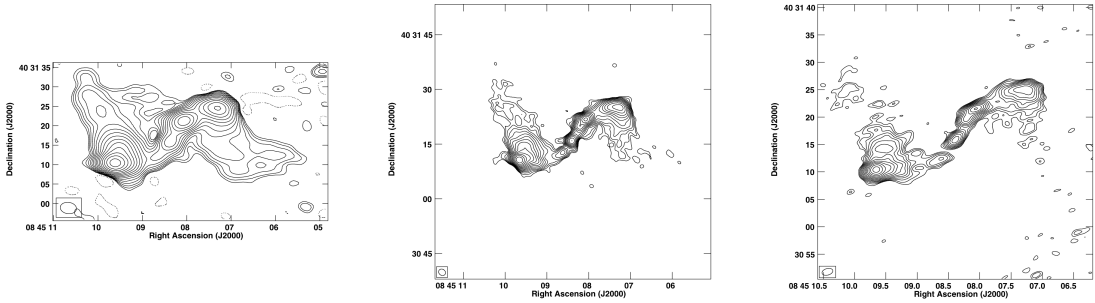

(d) J0845+4031, B-array L-band.(e) J0845+4031, B-array S-band.(f) J0845+4031, A-array L-band.

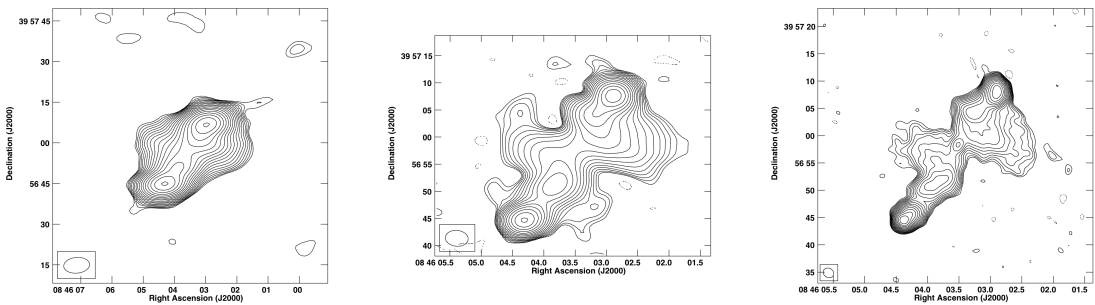

(g) J0846+3956, C-array S-band.(h) J0846+3956, B-array L-band.(i) J0846+3956, B-array S-band.

Fig. 9.—: VLA images of a sample of 100 low axial ratio radio sources (9/27).

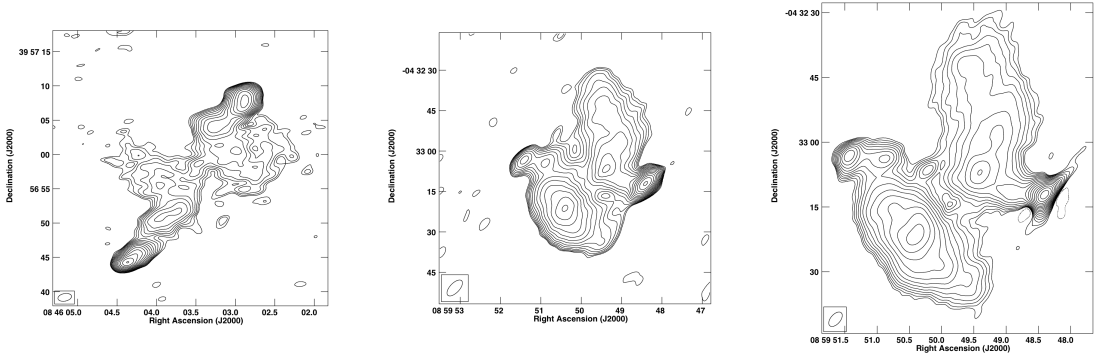

(a) J0846+3956, A-array L-band.(b) J0859−0433, B-array L-band.(c) J0859−0433, B-array S-band.

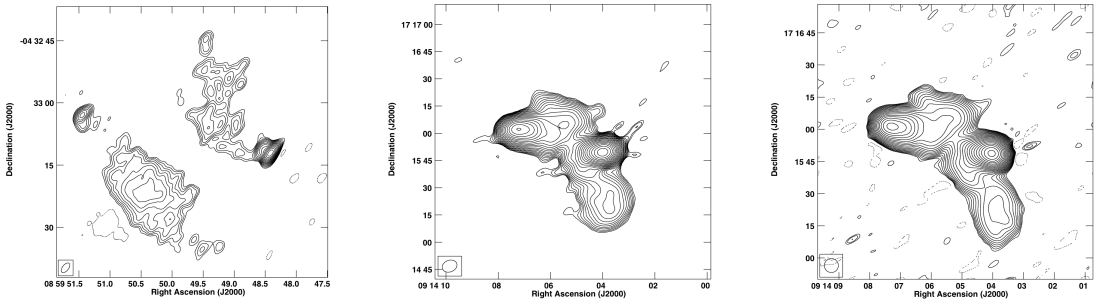

(d) J0859−0433, A-array L-band.(e) J0914+1715, C-array S-band.(f) J0914+1715, B-array L-band.

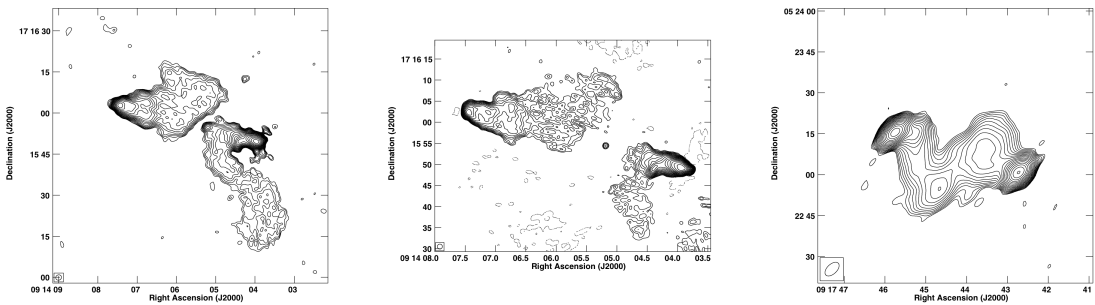

(g) J0914+1715, B-array S-band.(h) J0914+1715, A-array L-band.(i) J0917+0523, B-array L-band.

Fig. 10.—: VLA images of a sample of 100 low axial ratio radio sources (10/27).

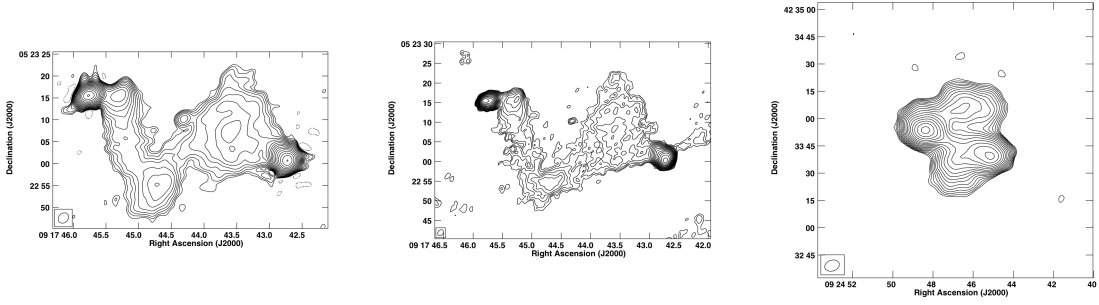

(a) J0917+0523, B-array S-band.(b) J0917+0523, A-array L-band.(c) J0924+4233, C-array S-band.

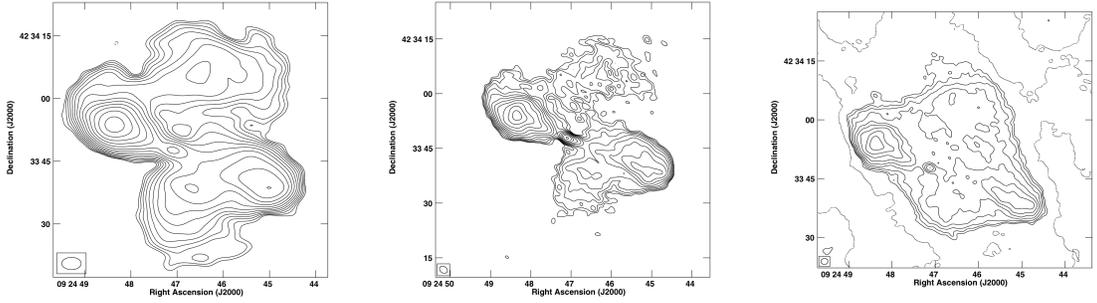

(d) J0924+4233, B-array L-band.(e) J0924+4233, B-array S-band.(f) J0924+4233, A-array L-band.

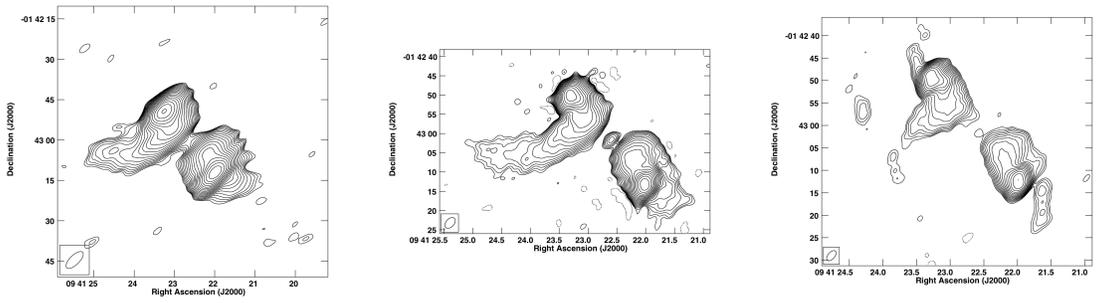

(g) J0941−0143, B-array L-band.(h) J0941−0143, B-array S-band.(i) J0941−0143, A-array L-band.

Fig. 11.—: VLA images of a sample of 100 low axial ratio radio sources (11/27).

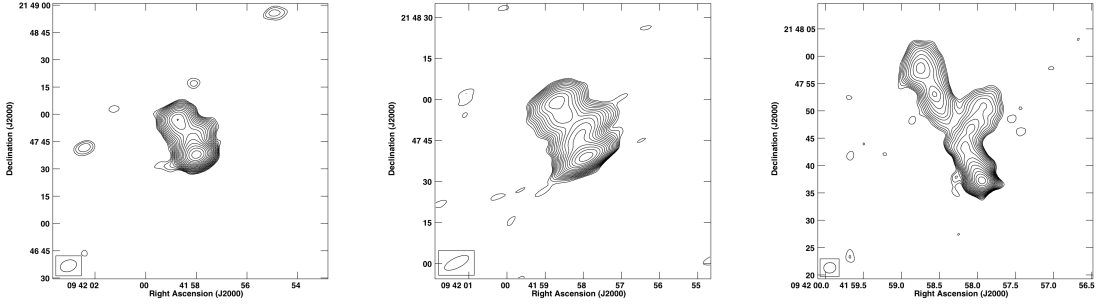

(a) J0941+2147, C-array S-band.(b) J0941+2147, B-array L-band.(c) J0941+2147, B-array S-band.

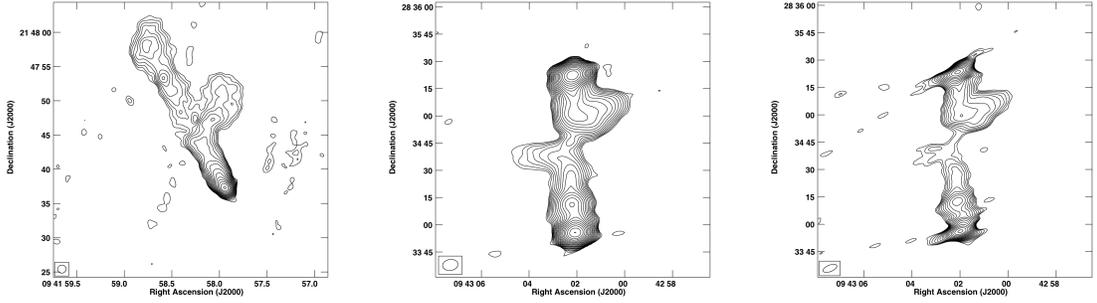

(d) J0941+2147, A-array L-band.(e) J0943+2834, C-array S-band.(f) J0943+2834, B-array L-band.

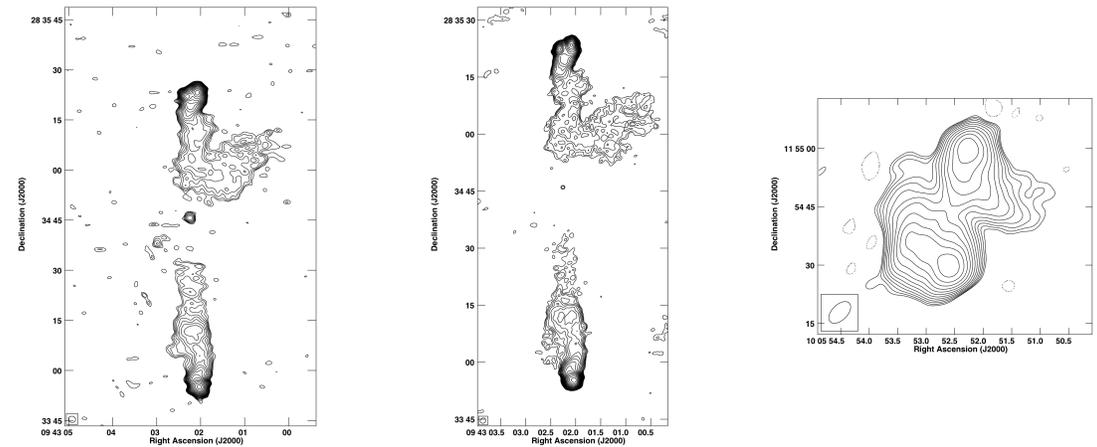

(g) J0943+2834, B-array S-band.(h) J0943+2834, A-array L-band.(i) J1005+1154, B-array L-band.

Fig. 12.—: VLA images of a sample of 100 low axial ratio radio sources (12/27).

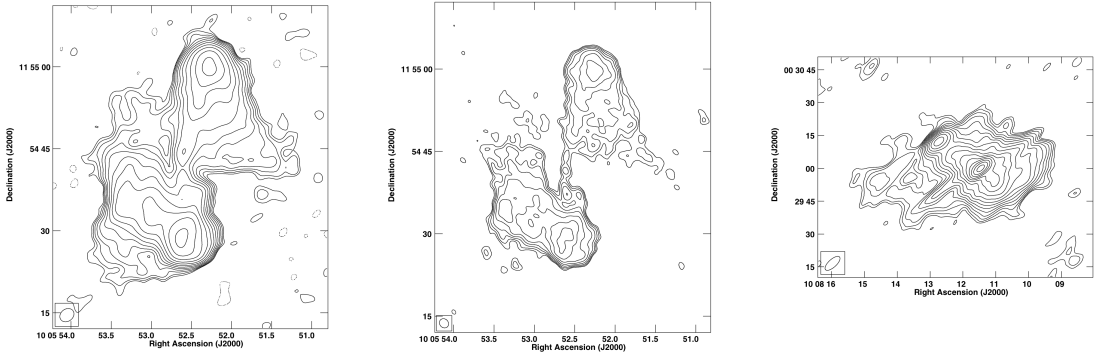

(a) J1005+1154, B-array S-band.(b) J1005+1154, A-array L-band.(c) J1008+0030, B-array L-band.

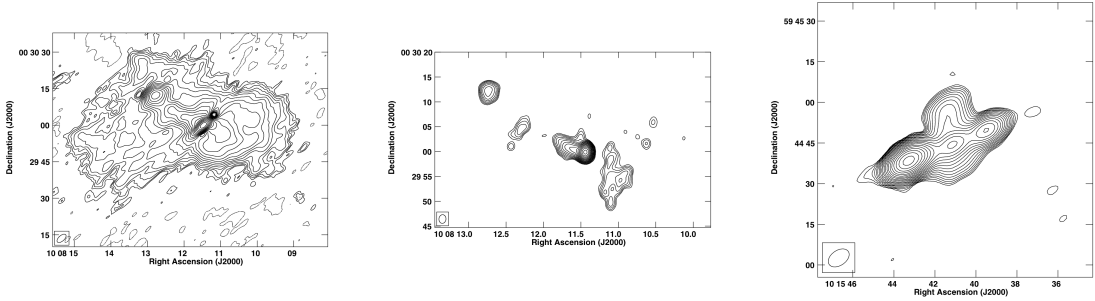

(d) J1008+0030, B-array S-band.(e) J1008+0030, A-array L-band.(f) J1015+5944, C-array S-band.

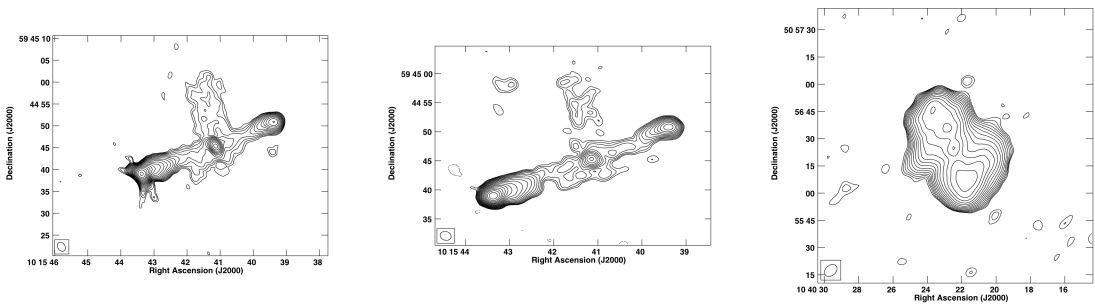

(g) J1015+5944, B-array S-band.(h) J1015+5944, A-array L-band.(i) J1040+5056, C-array S-band.

Fig. 13.—: VLA images of a sample of 100 low axial ratio radio sources (13/27).

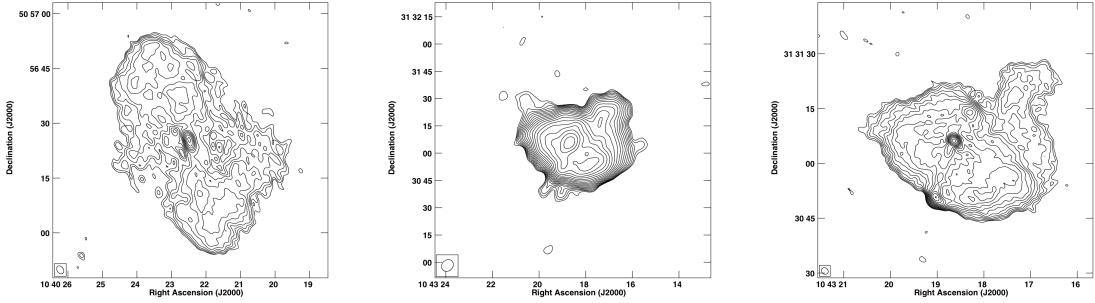

(a) J1040+5056, B-array S-band.(b) J1043+3131, C-array S-band.(c) J1043+3131, B-array S-band.

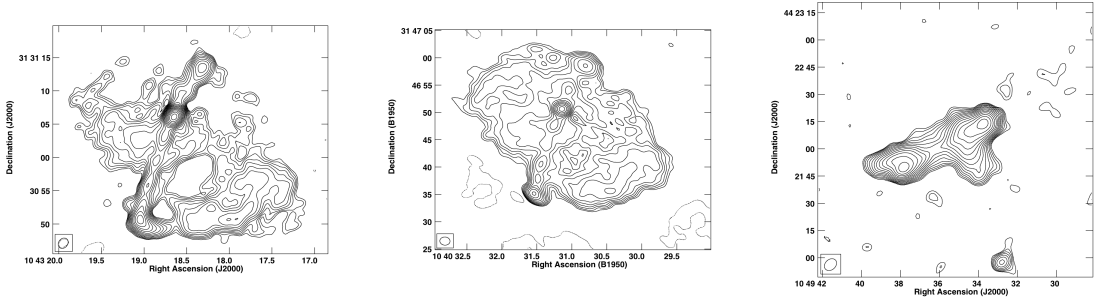

(d) J1043+3131, B-array C-band.(e) J1043+3131, A-array L-band.(f) J1049+4422, C-array S-band.

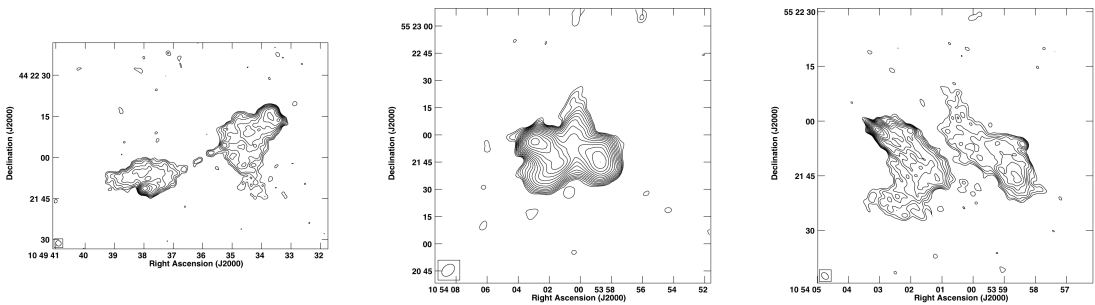

(g) J1049+4422, B-array S-band.(h) J1054+5521, C-array S-band.(i) J1054+5521, B-array S-band.

Fig. 14.—: VLA images of a sample of 100 low axial ratio radio sources (14/27).

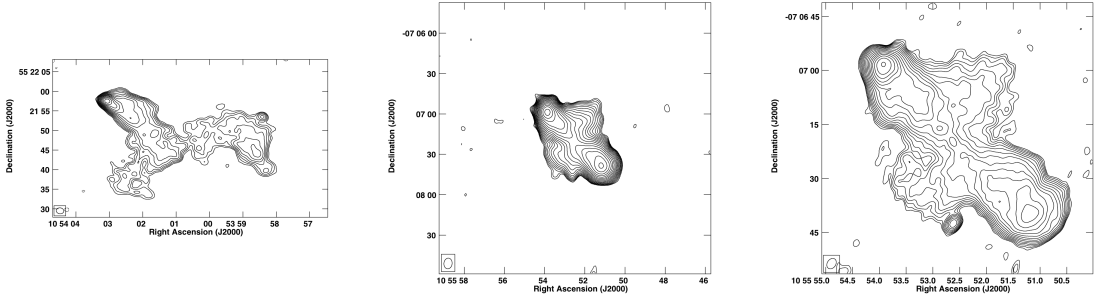

(a) J1054+5521, A-array L-band.(b) J1055−0707, C-array S-band.(c) J1055−0707, B-array S-band.

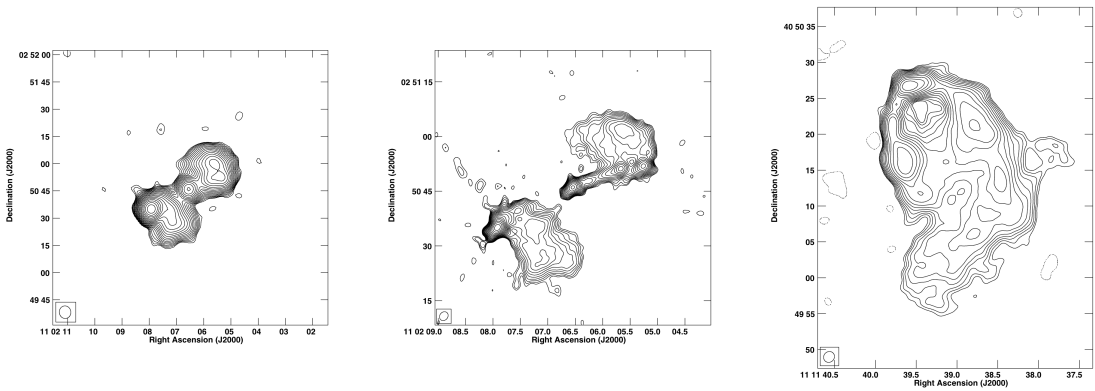

(d) J1102+0250, C-array S-band.(e) J1102+0250, B-array S-band.(f) J1111+4050, B-array C-band.

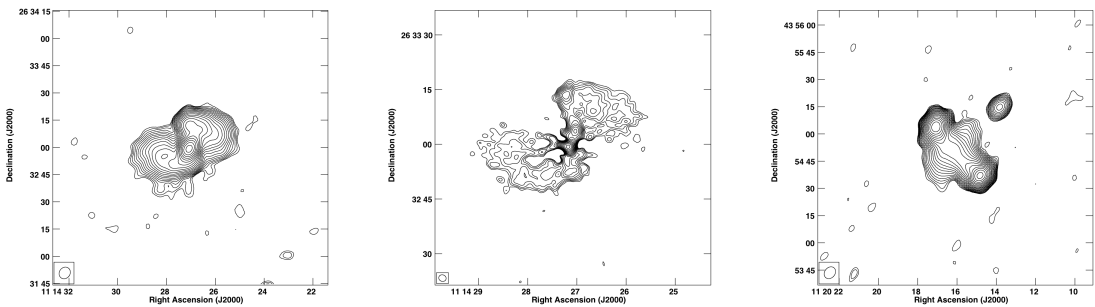

(g) J1114+2632, C-array S-band.(h) J1114+2632, B-array S-band.(i) J1120+4354, C-array S-band.

Fig. 15.—: VLA images of a sample of 100 low axial ratio radio sources (15/27).

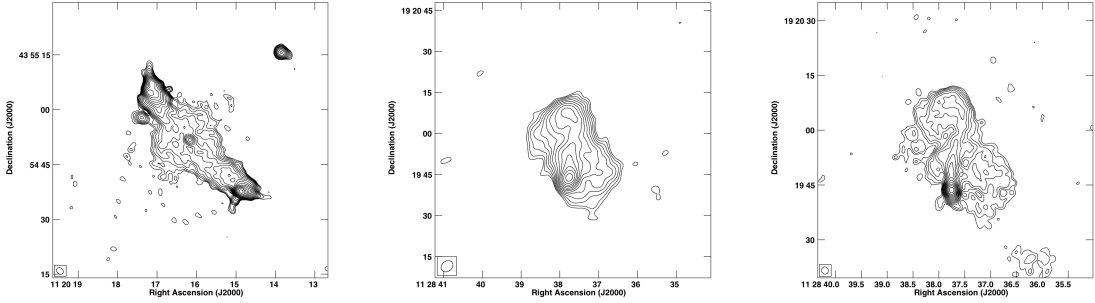

(a) J1120+4354, B-array S-band. (b) J1128+1919, B-array L-band. (c) J1128+1919, B-array S-band.

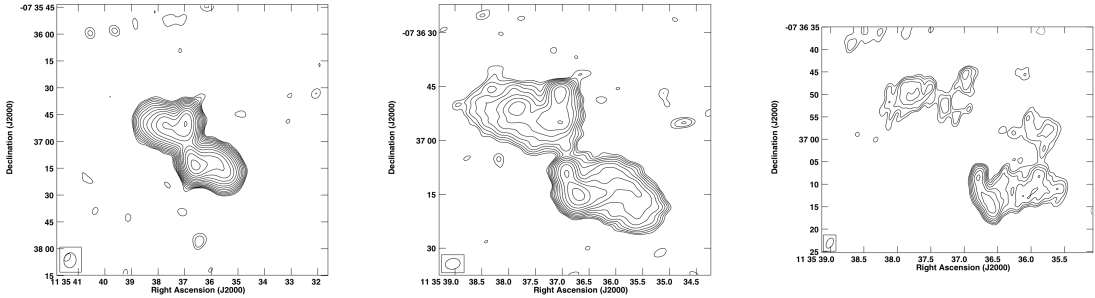

(d) J1135−0737, C-array S-band. (e) J1135−0737, B-array S-band. (f) J1135−0737, A-array L-band.

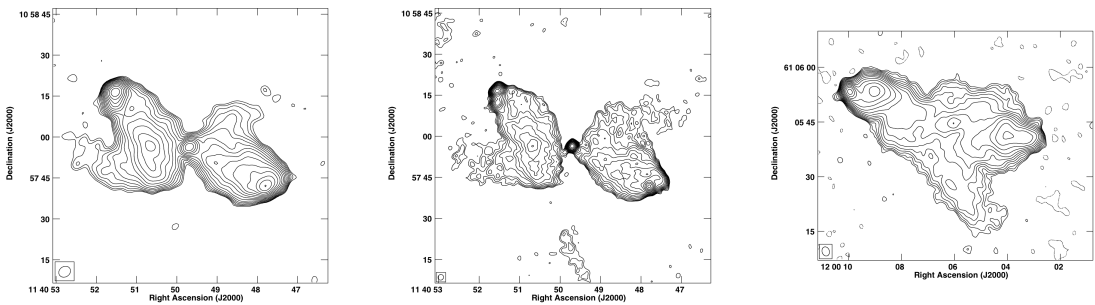

(g) J1140+1057, B-array L-band. (h) J1140+1057, B-array S-band. (i) J1200+6105, B-array S-band.

Fig. 16.—: VLA images of a sample of 100 low axial ratio radio sources (16/27).

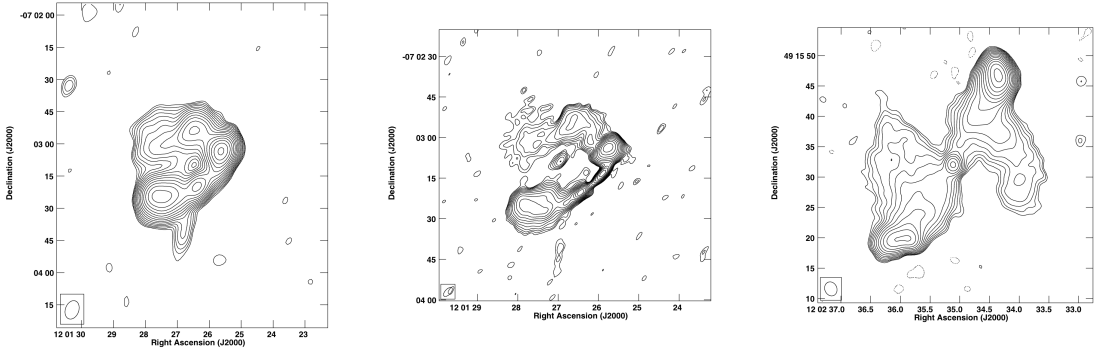

(a) J1201−0703, C-array S-band.(b) J1201−0703, B-array S-band.(c) J1202+4915, B-array S-band.

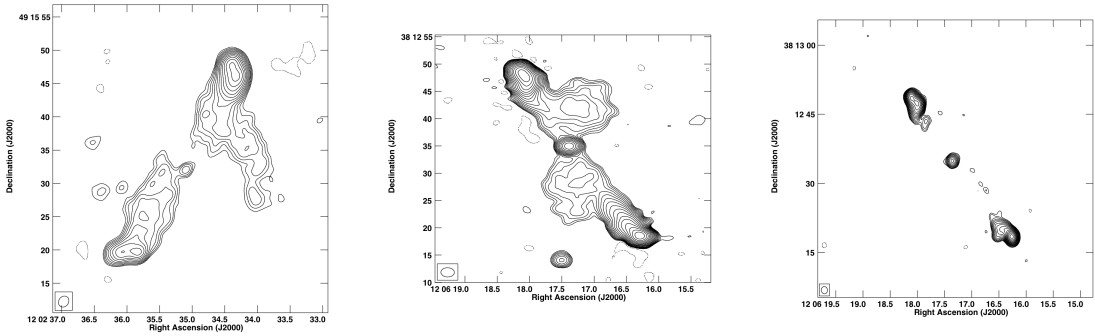

(d) J1202+4915, A-array L-band.(e) J1206+3812, B-array S-band.(f) J1206+3812, B-array C-band.

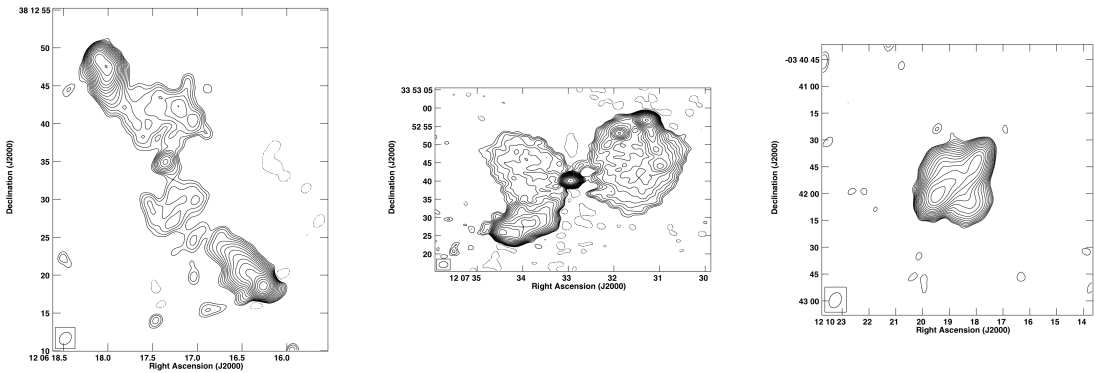

(g) J1206+3812, A-array L-band.(h) J1207+3352, B-array S-band.(i) J1210−0341, C-array S-band.

Fig. 17.—: VLA images of a sample of 100 low axial ratio radio sources (17/27).

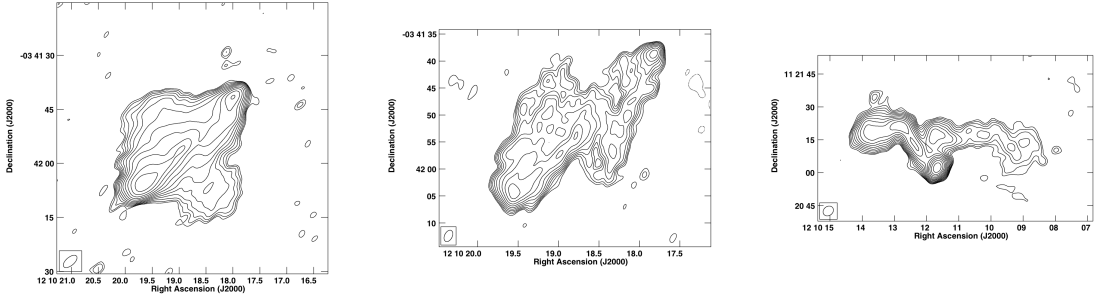

(a) J1210−0341, B-array S-band.(b) J1210−0341, A-array L-band.(c) J1210+1121, B-array L-band.

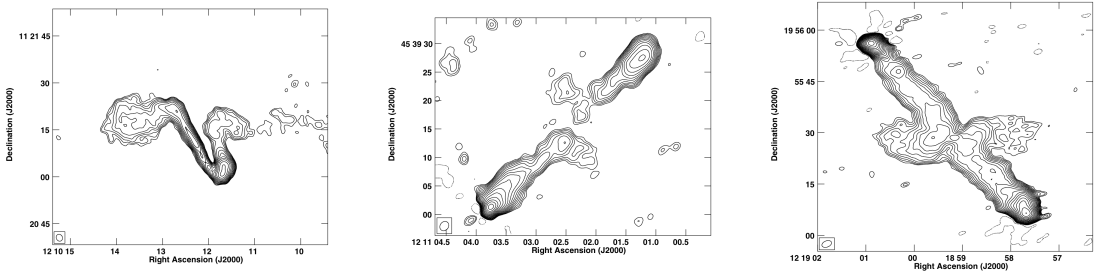

(d) J1210+1121, B-array S-band.(e) J1211+4539, A-array L-band.(f) J1218+1955, B-array S-band.

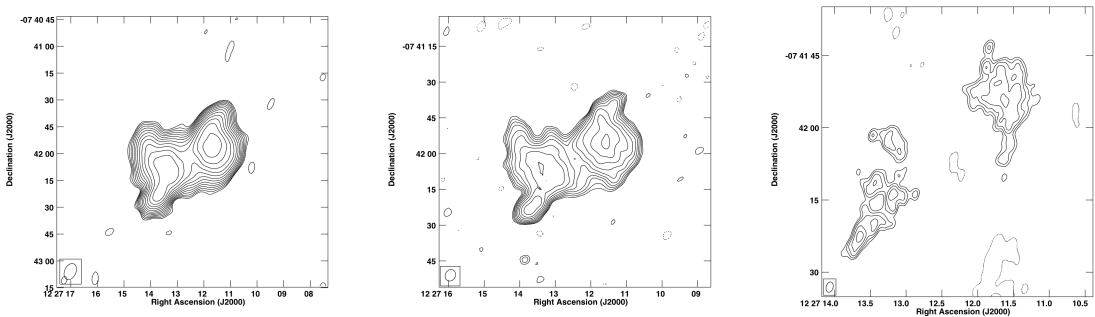

(g) J1227−0742, C-array S-band.(h) J1227−0742, B-array S-band.(i) J1227−0742, A-array L-band.

Fig. 18.—: VLA images of a sample of 100 low axial ratio radio sources (18/27).

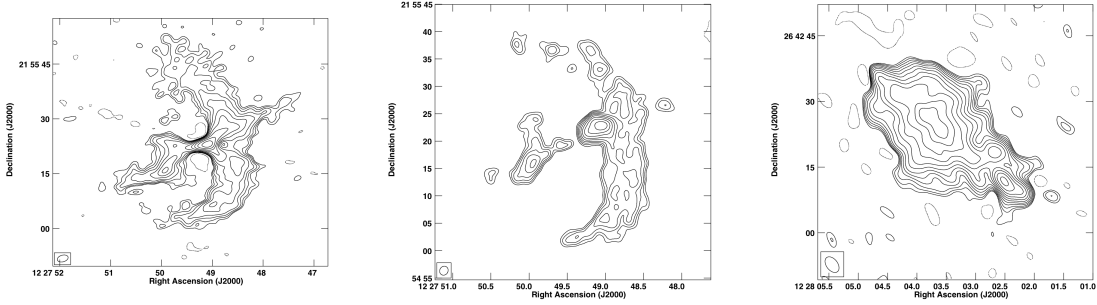

(a) J1227+2155, B-array S-band.(b) J1227+2155, A-array L-band.(c) J1228+2642, B-array S-band.

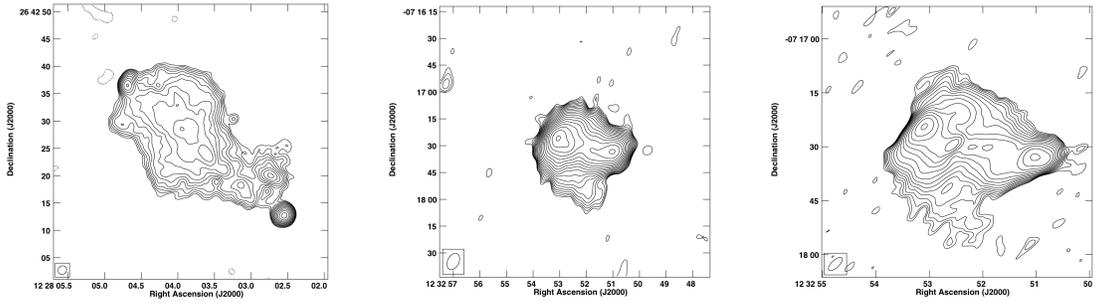

(d) J1228+2642, A-array L-band.(e) J1232−0717, C-array S-band.(f) J1232−0717, B-array S-band.

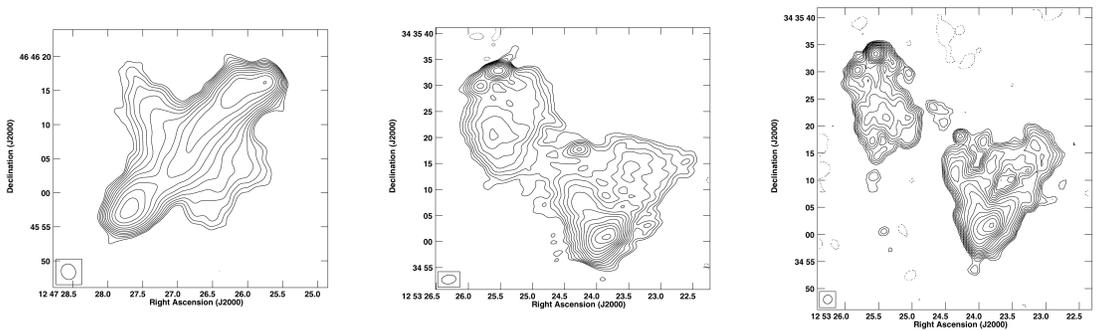

(g) J1247+4646, B-array S-band.(h) J1253+3435, B-array S-band.(i) J1253+3435, A-array L-band.

Fig. 19.—: VLA images of a sample of 100 low axial ratio radio sources (19/27).

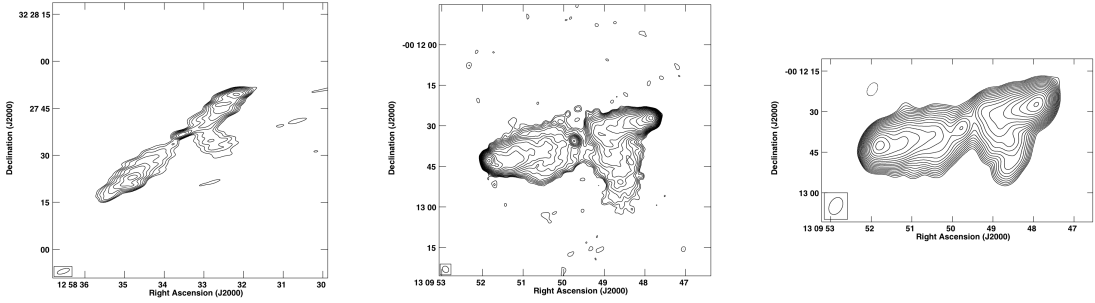

(a) J1258+3227, B-array S-band.(b) J1309−0012, B-array S-band.(c) J1309−0012, B-array C-band.

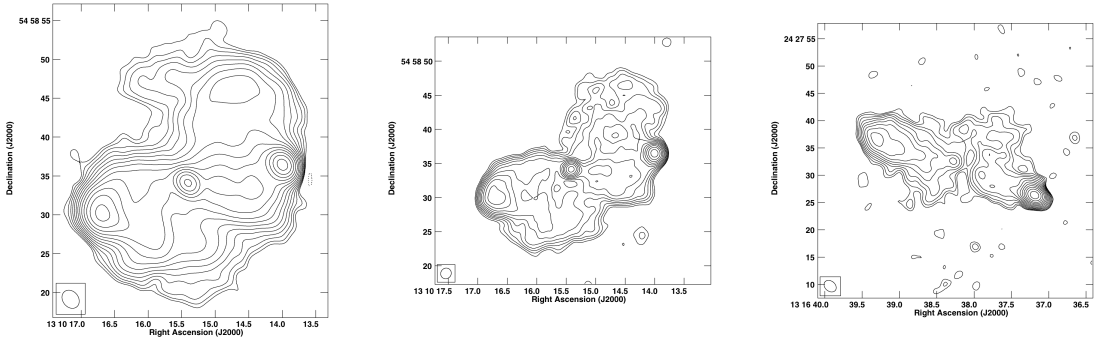

(d) J1310+5458, B-array S-band.(e) J1310+5458, A-array L-band.(f) J1316+2427, B-array S-band.

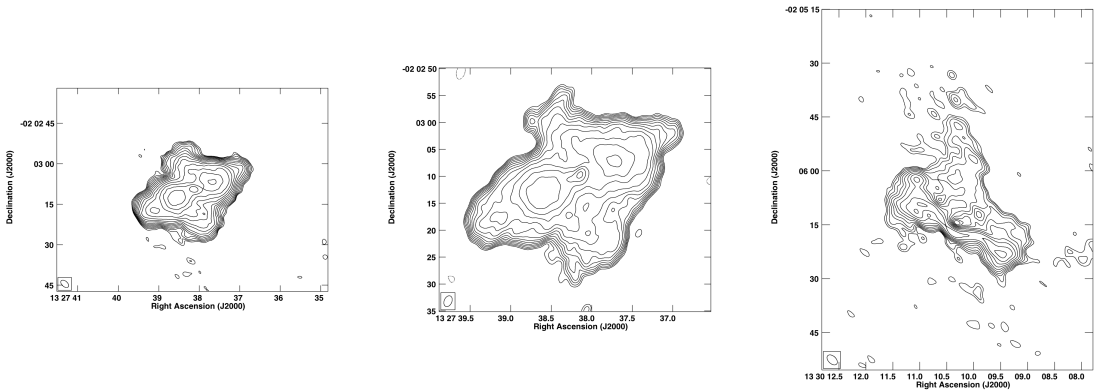

(g) J1327−0203, B-array S-band.(h) J1327−0203, A-array L-band.(i) J1330−0206, B-array S-band.

Fig. 20.—: VLA images of a sample of 100 low axial ratio radio sources (20/27).

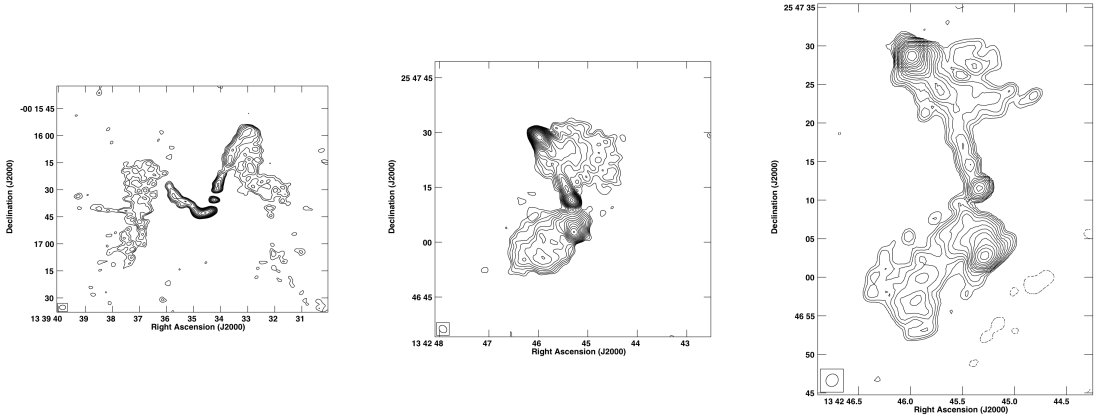

(a) J1339−0016, B-array S-band.(b) J1342+2547, B-array S-band.(c) J1342+2547, A-array L-band.

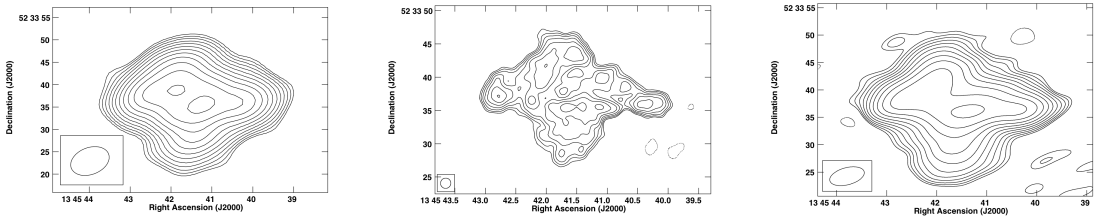

(d) J1345+5233, C-array S-band.(e) J1345+5233, A-array L-band.(f)   J1345+5233,   AB-array   S-band.

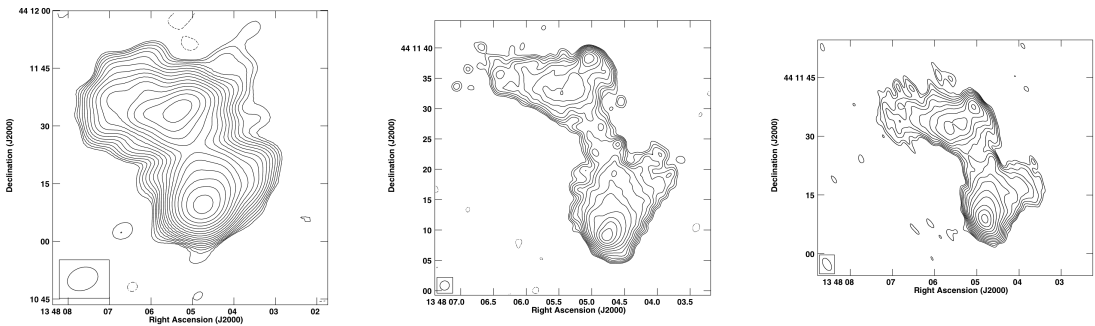

(g) J1348+4411, C-array S-band.(h) J1348+4411, A-array L-band.(i)   J1348+4411,   AB-array   S-band.

Fig. 31.— VLA images for a sample of 100 low-axial-ratio radio sources (31/37).

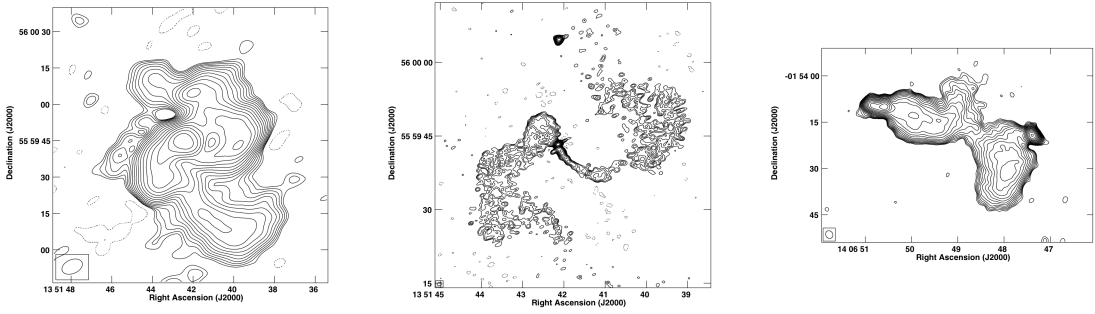

(a) J1351+5559, C-array S-band. (b) J1351+5559, AB-array S-band. (c) J1406−0154, B-array S-band.

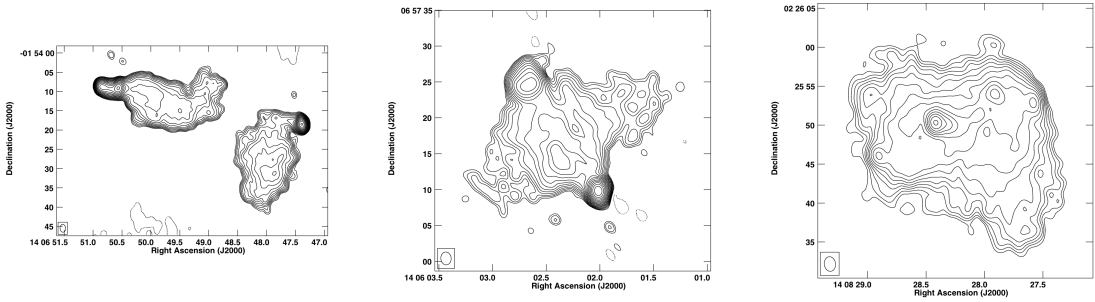

(d) J1406−0154, A-array L-band. (e) J1406+0657, A-array L-band. (f) J1408+0225, A-array L-band.

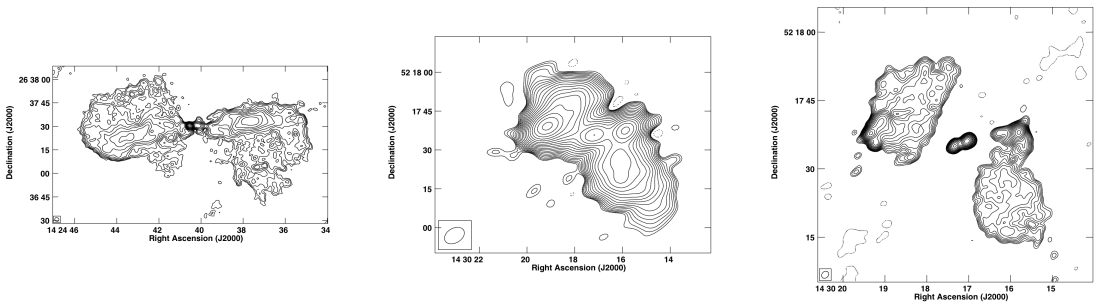

(g) J1424+2637, B-array S-band. (h) J1430+5217, C-array S-band. (i) J1430+5217, A-array L-band.

Fig. 22.—: VLA images of a sample of 100 low axial ratio radio sources (22/27).

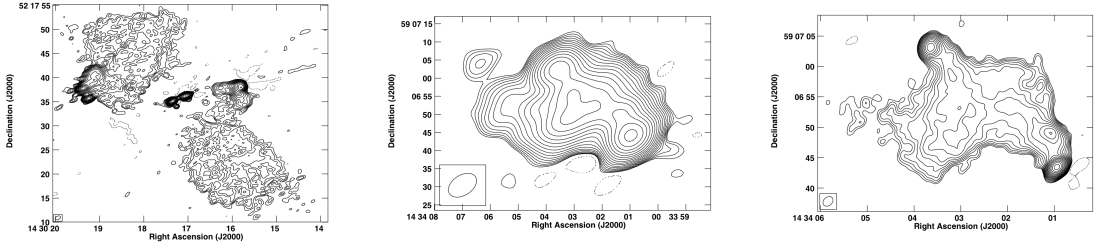

(a) J1430+5217, AB-array S-(b) J1434+5906, C-array S-band.(c) J1434+5906, A-array L-band.
band.

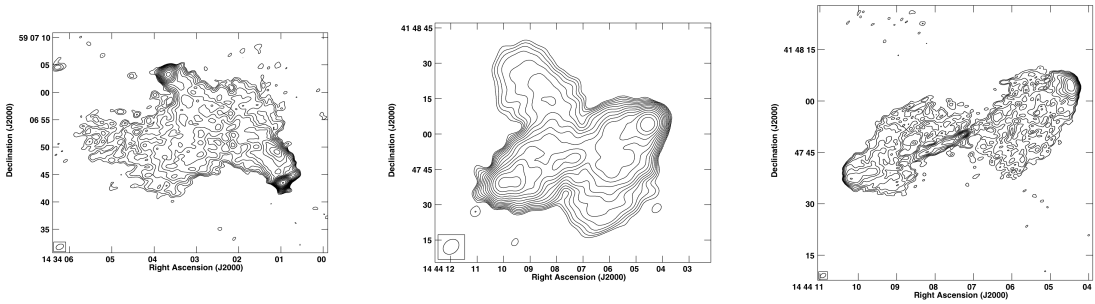

(d) J1434+5906, AB-array S-(e) J1444+4147, C-array S-band.(f) J1444+4147, AB-array S-
band. band.

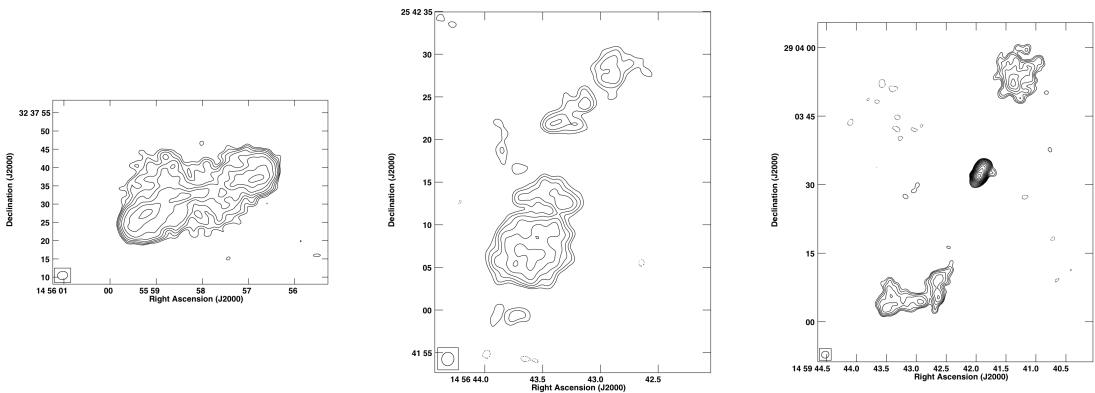

(g) J1455+3237, B-array S-band.(h) J1456+2542, A-array L-band.(i) J1459+2903, B-array C-band.

Fig. 23.— VLA images for sample of 100 low-redshift radio sources (23/27).

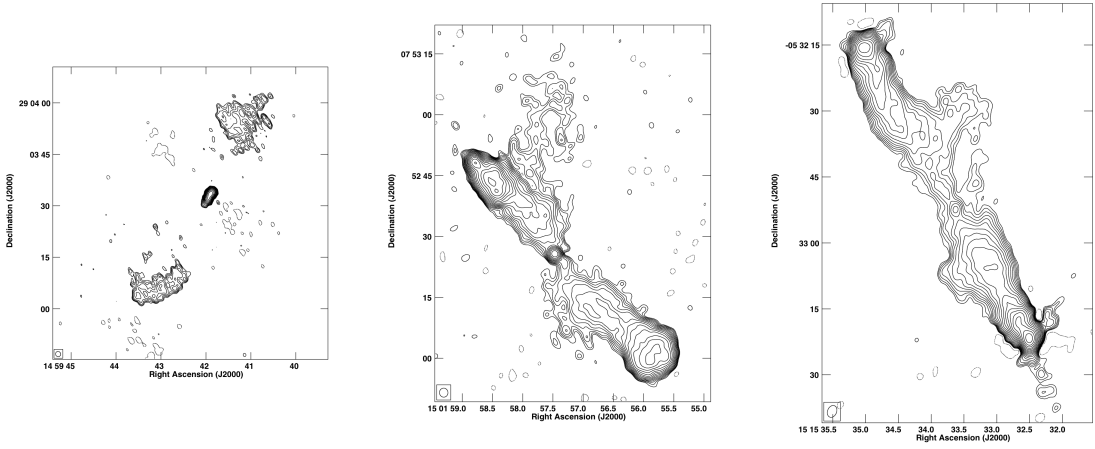

(a) J1459+2903, A-array L-band.(b) J1501+0752, B-array S-band.(c) J1515−0532, B-array S-band.

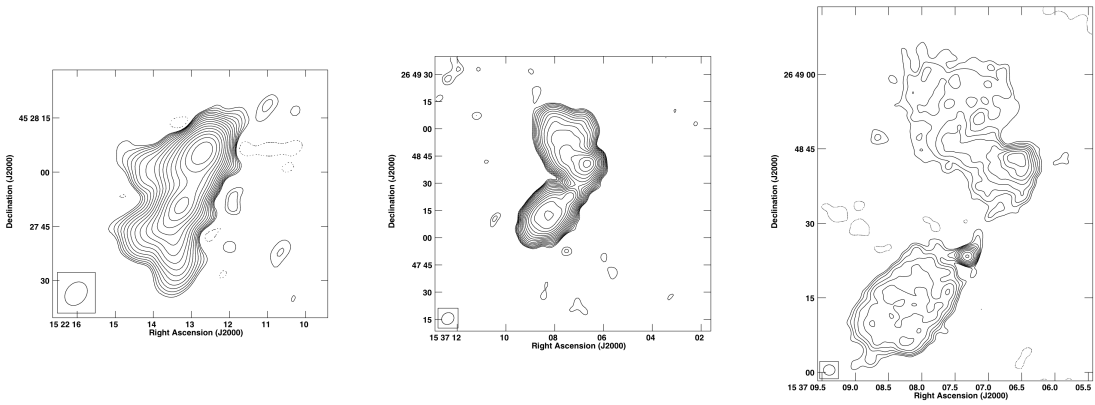

(d) J1522+4527, C-array S-band.(e) J1537+2648, C-array S-band.(f) J1537+2648, B-array S-band.

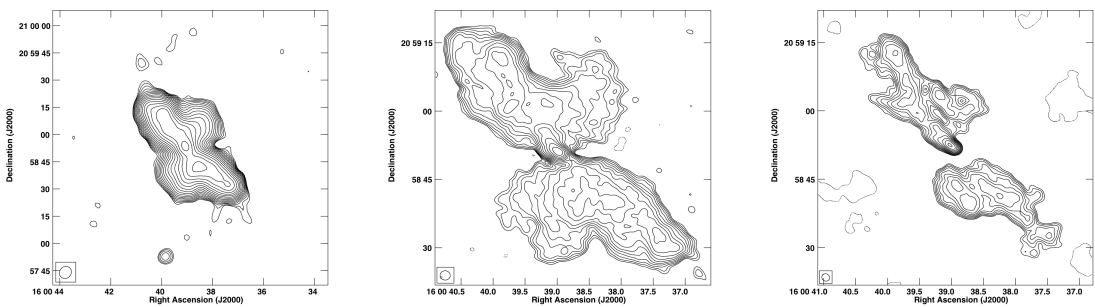

(g) J1600+2058, C-array S-band.(h) J1600+2058, B-array S-band.(i) J1600+2058, A-array L-band.

Fig. 24.—: VLA images of a sample of 100 low axial ratio radio sources (24/27).

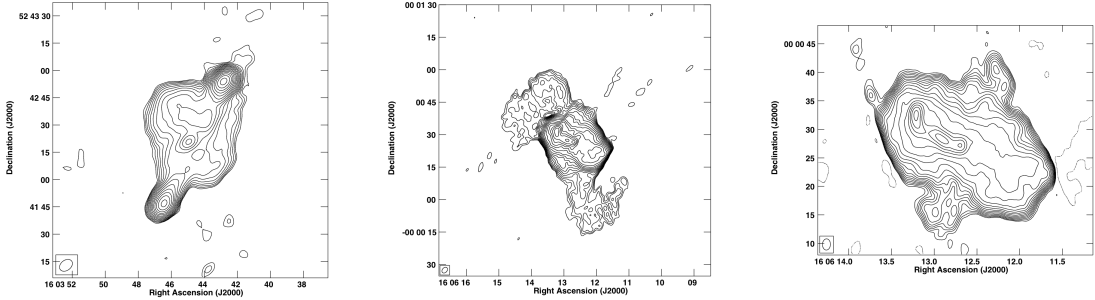

(a) J1603+5242, C-array S-band.(b) J1606+0000, B-array S-band.(c) J1606+0000, A-array L-band.

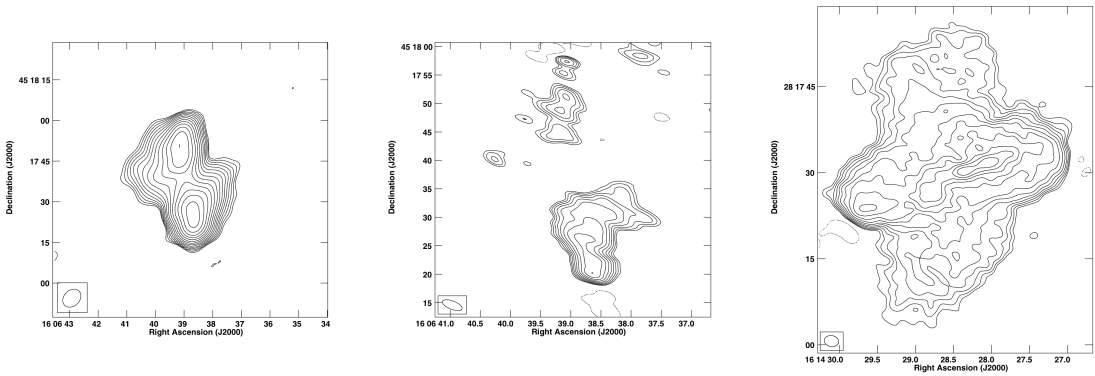

(d) J1606+4517, C-array S-band.(e) J1606+4517, A-array L-band.(f) J1614+2817, B-array S-band.

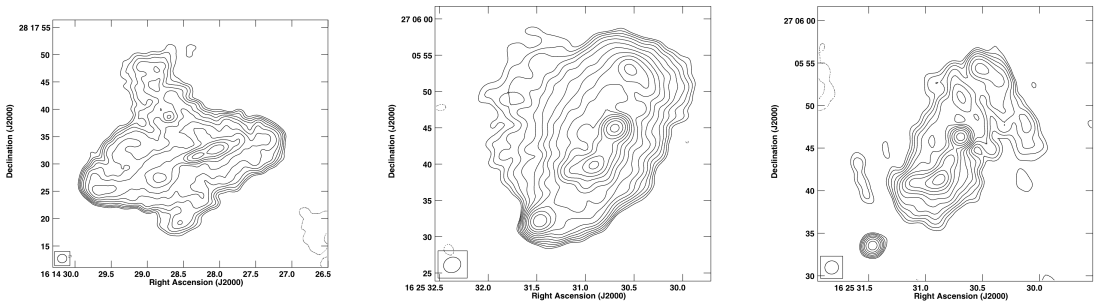

(g) J1614+2817, A-array L-band.(h) J1625+2705, B-array S-band.(i) J1625+2705, A-array L-band.

Fig. 25.—: VLA images of a sample of 100 low axial ratio radio sources (25/27).

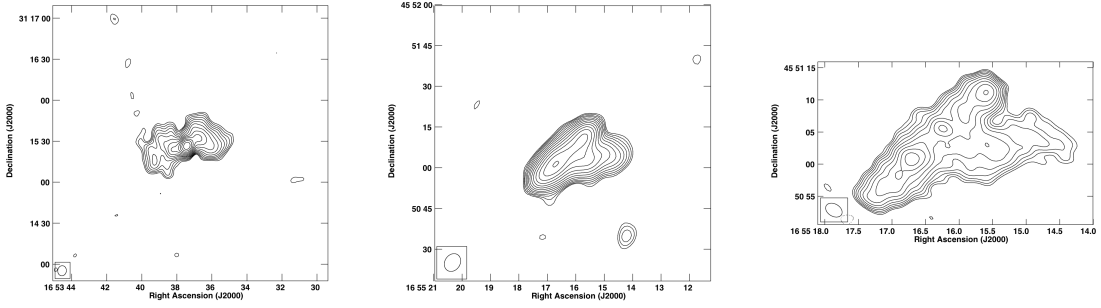

(a) J1653+3115, C-array S-band.(b) J1655+4551, C-array S-band.(c) J1655+4551, B-array S-band.

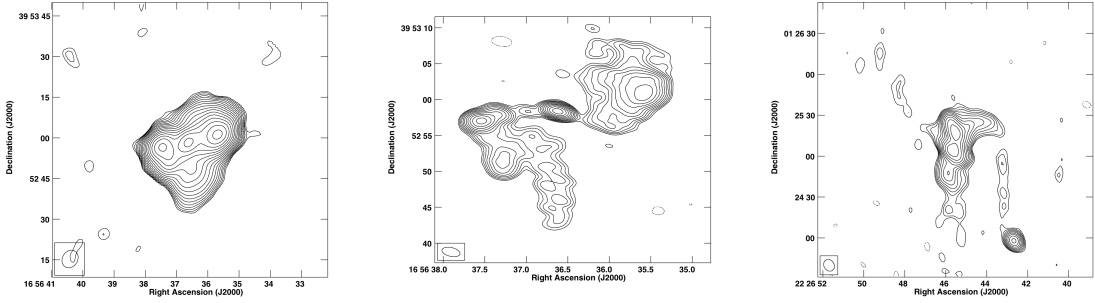

(d) J1656+3952, C-array S-band.(e) J1656+3952, A-array L-band.(f) J2226+0125, C-array S-band.

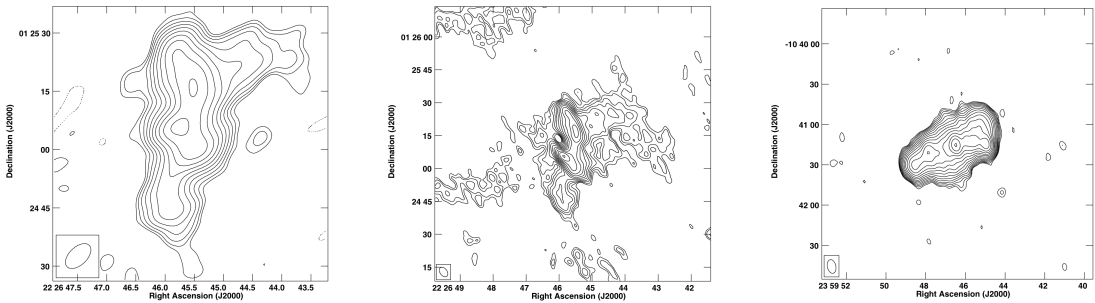

(g) J2226+0125, B-array L-band.(h) J2226+0125, B-array S-band.(i) J2359−1041, C-array S-band.

Fig. 26.—: VLA images of a sample of 100 low axial ratio radio sources (26/27).

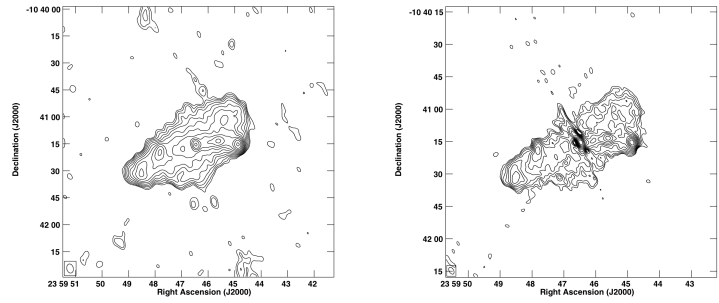

(a) J2359−1041, B-array L-band.(b) J2359−1041, B-array S-band.

Fig. 27.—: VLA images of a sample of 100 low axial ratio radio sources (27/27).

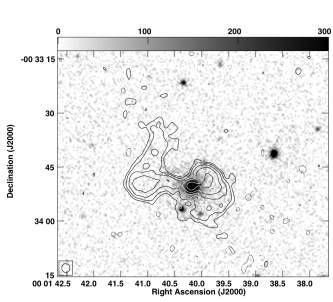 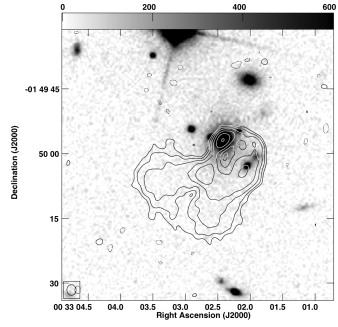 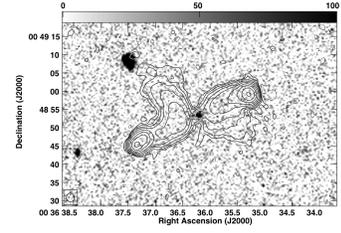

(a) J0001−0033          (b) J0033−0149          (c) J0036+0048

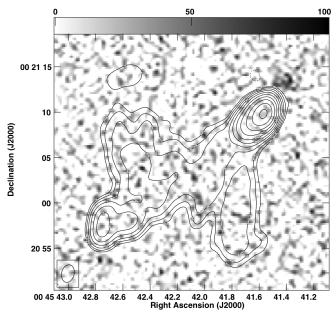 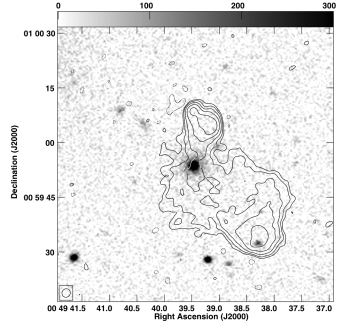 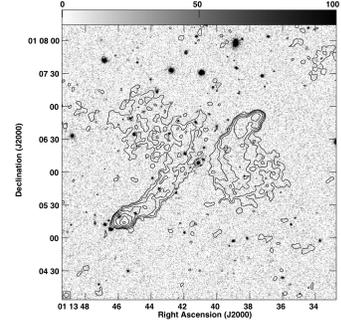

(d) J0045+0021          (e) J0049+0059          (f) J0113+0106

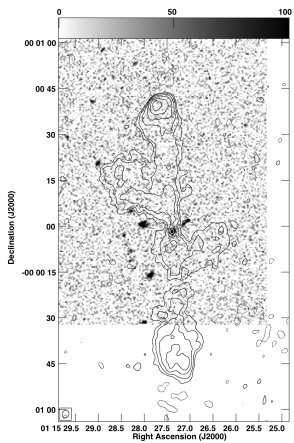 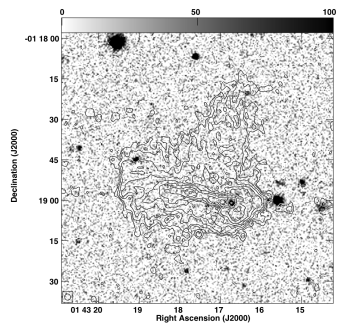 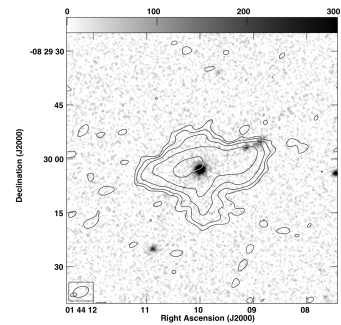

(g) J0115−0000          (h) J0143−0119          (i) J0144−0830

Fig. 28.—: Optical overlays of a sample of 100 low axial ratio radio sources (1/11).

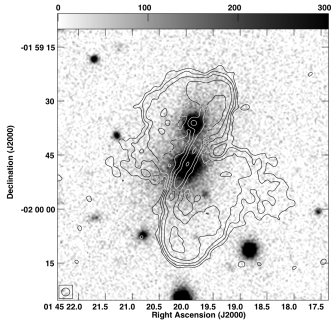

(a) J0145−0159

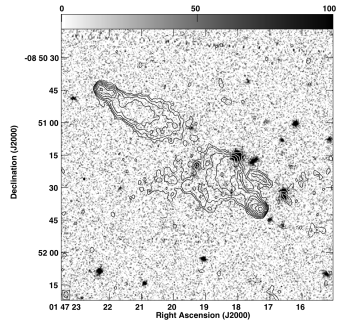

(b) J0147−0851

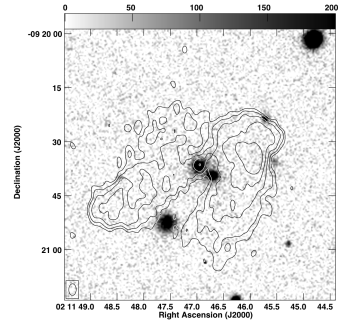

(c) J0211−0920

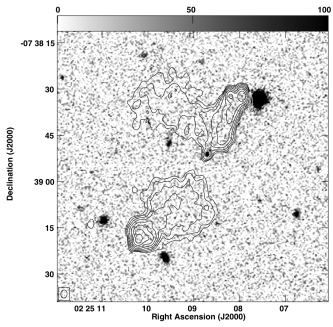

(d) J0225−0738

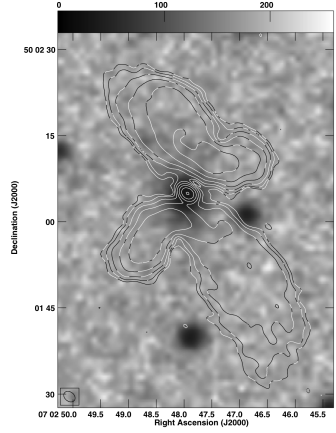

(e) J0702+5002

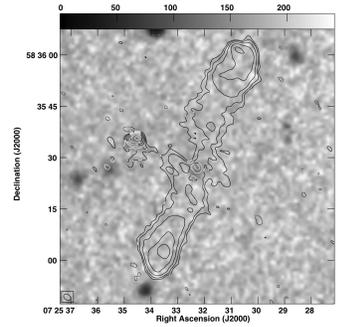

(f) J0725+5835

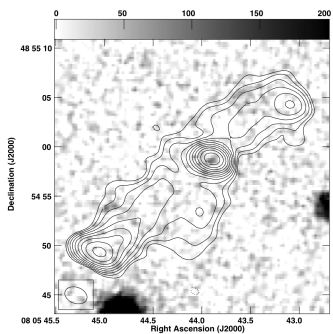

(g) J0805+4854

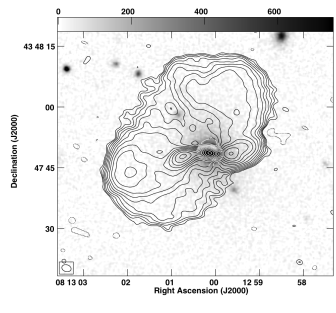

(h) J0813+4347

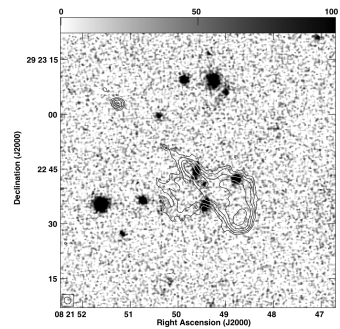

(i) J0821+2922

Fig. 29.—: Optical overlays of a sample of 100 low axial ratio radio sources (2/11).

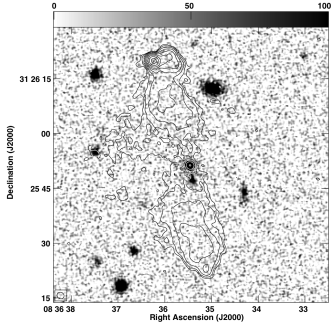

(a) J0836+3125

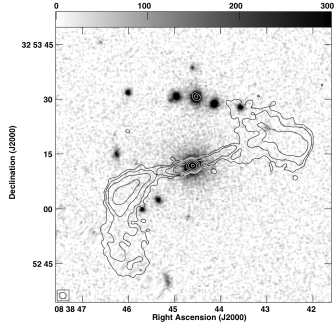

(b) J0838+3253

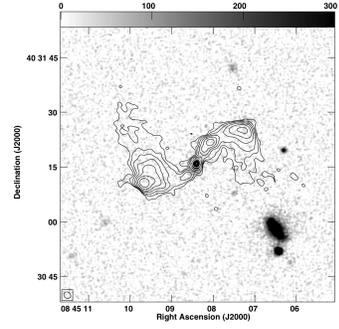

(c) J0845+4031

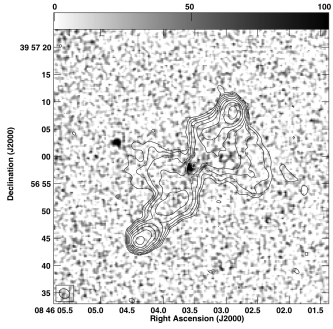

(d) J0846+3956

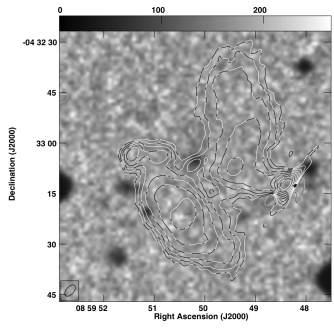

(e) J0859−0433

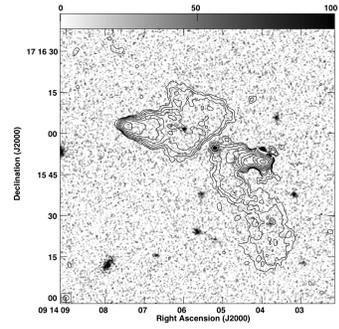

(f) J0914+1715

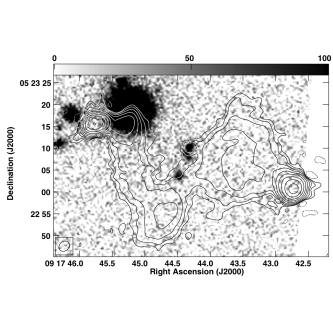

(g) J0917+0523

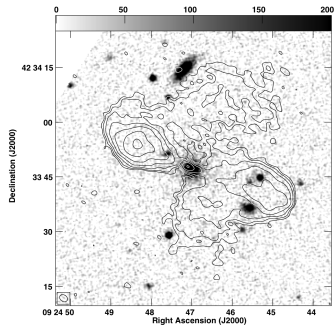

(h) J0924+4233

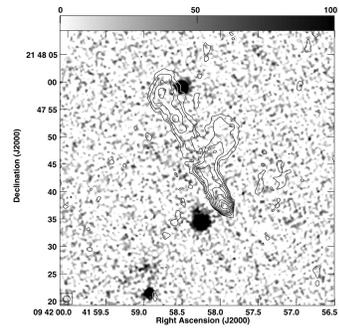

(i) J0941+2147

Fig. 30.—: Optical overlays of a sample of 100 low axial ratio radio sources (3/11).

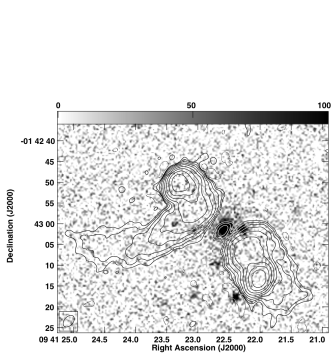

(a) J0941−0143

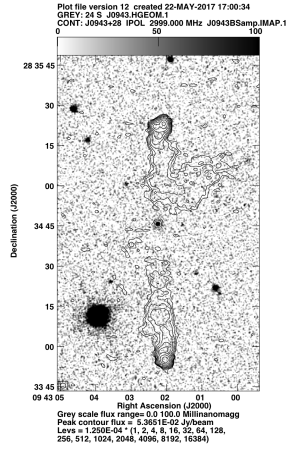

(b) J0943+2834

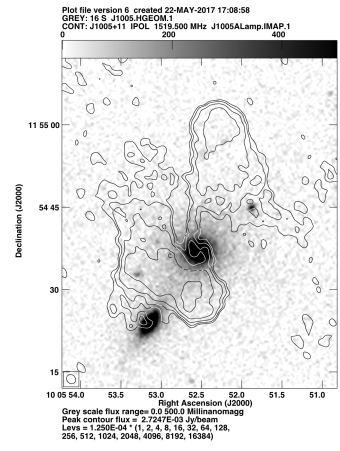

(c) J1005+1154

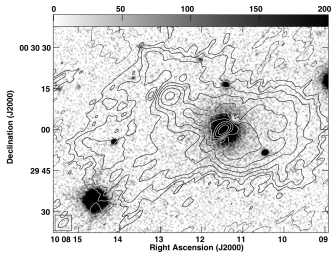

(d) J1008+0030

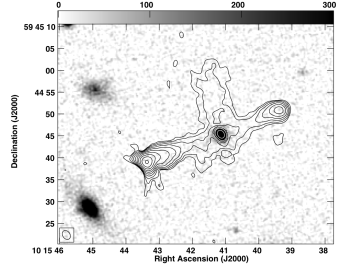

(e) J1015+5944

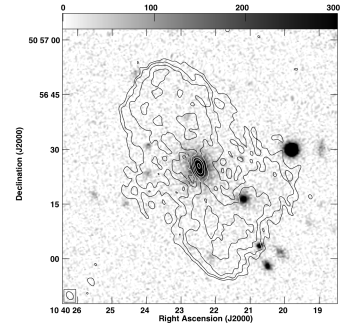

(f) J1040+5056

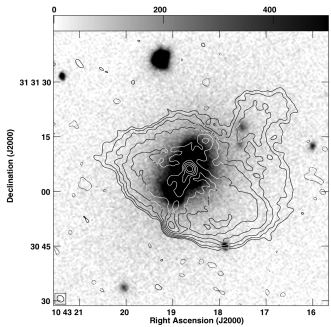

(g) J1043+3131

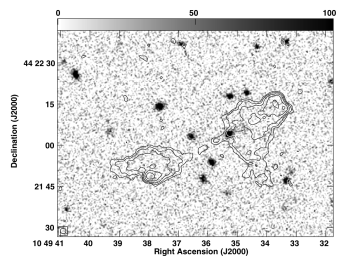

(h) J1049+4422

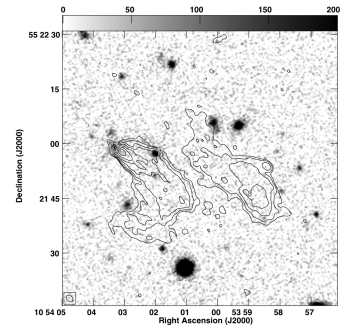

(i) J1054+5521

Fig. 31.—: Optical overlays of a sample of 100 low axial ratio radio sources (4/11).

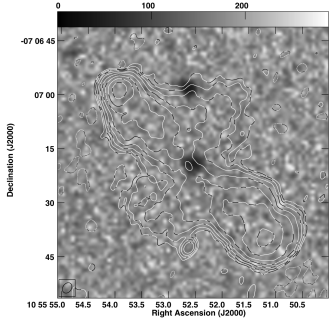

(a) J1055−0707

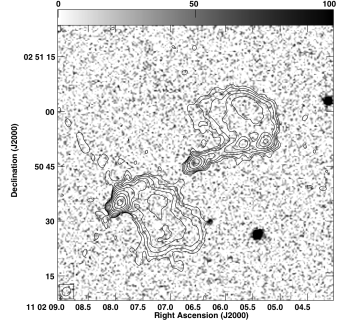

(b) J1102+0250

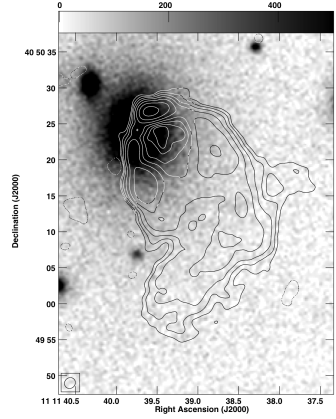

(c) J1111+4050

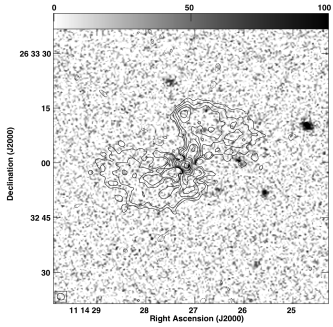

(d) J1114+2632

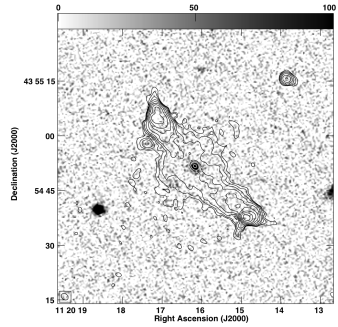

(e) J1120+4354

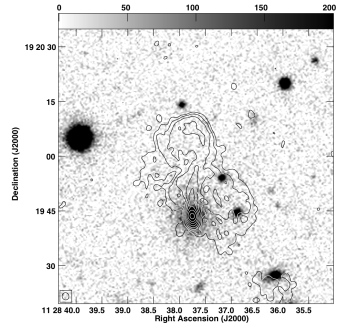

(f) J1128+1919

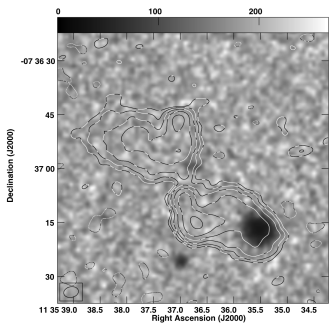

(g) J1135−0737

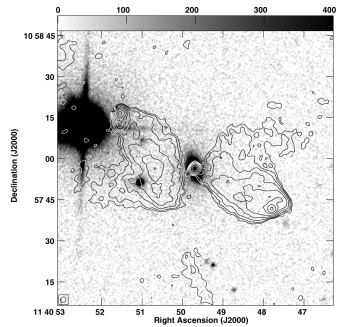

(h) J1140+1057

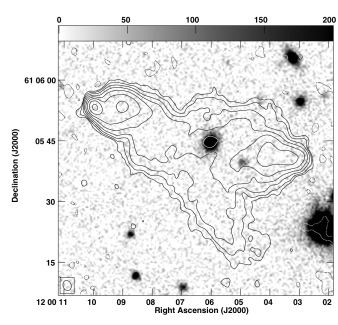

(i) J1200+6105

Fig. 32.—: Optical overlays of a sample of 100 low axial ratio radio sources (5/11).

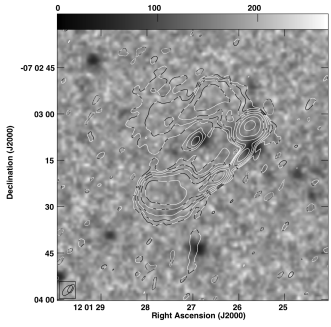

(a) J1201−0703

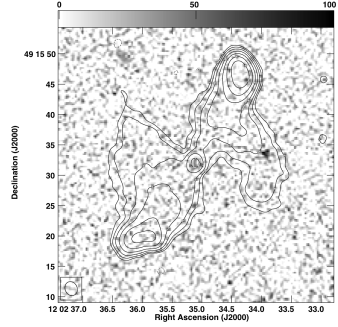

(b) J1202+4915

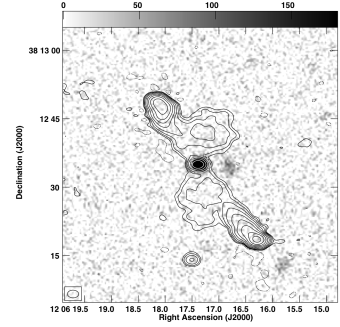

(c) J1206+3812

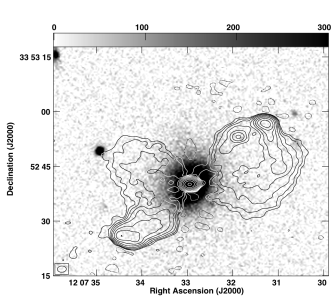

(d) J1207+3352

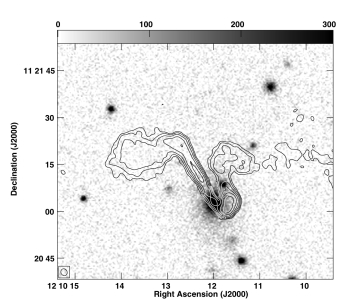

(e) J1210+1121

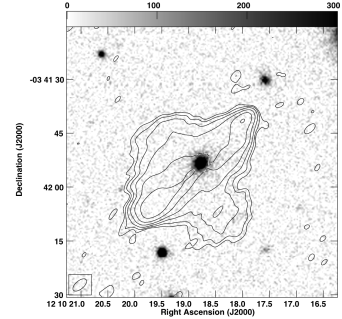

(f) J1210−0341

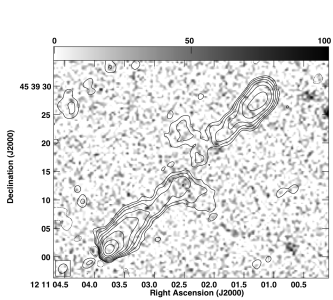

(g) J1211+4539

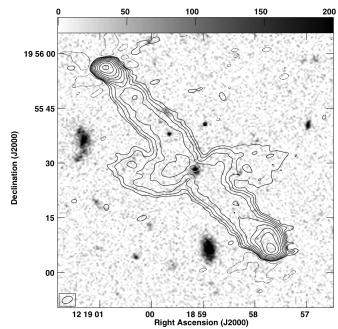

(h) J1218+1955

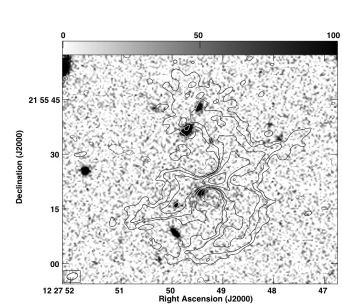

(i) J1227+2155

Fig. 33.—: Optical overlays of a sample of 100 low axial ratio radio sources (6/11).

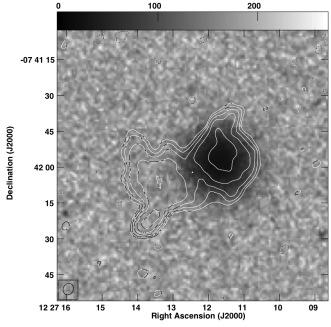

(a) J1227−0742

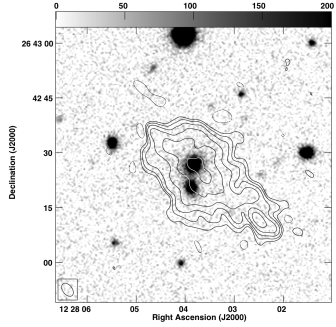

(b) J1228+2642

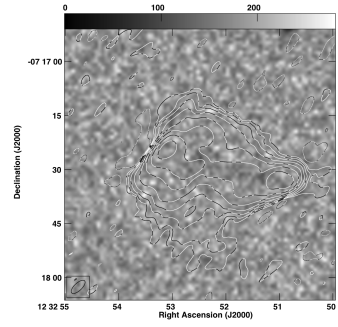

(c) J1232−0717

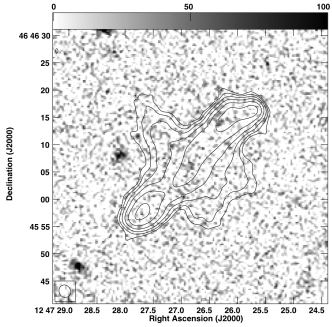

(d) J1247+4646

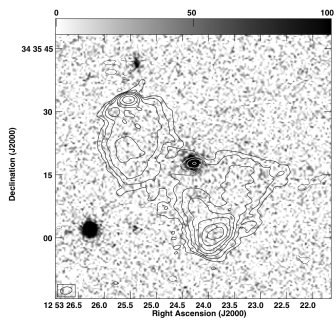

(e) J1253+3435

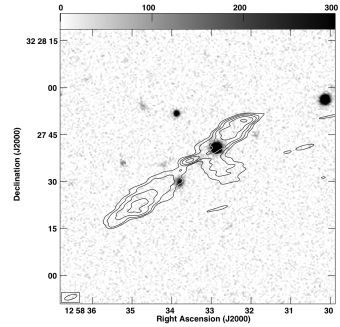

(f) J1258+3227

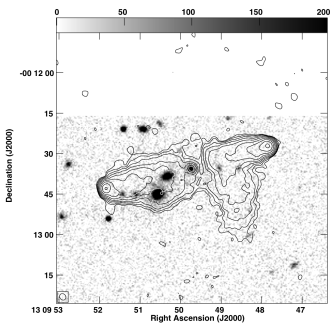

(g) J1309−0012

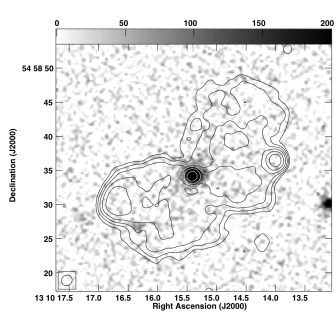

(h) J1310+5458

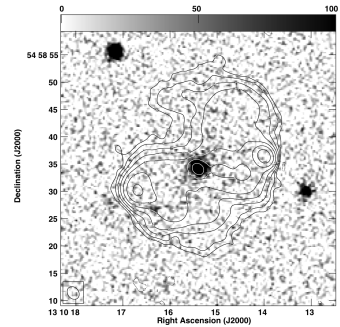

(i) J1310+5458

Fig. 34.—: Optical overlays of a sample of 100 low axial ratio radio sources (7/11).

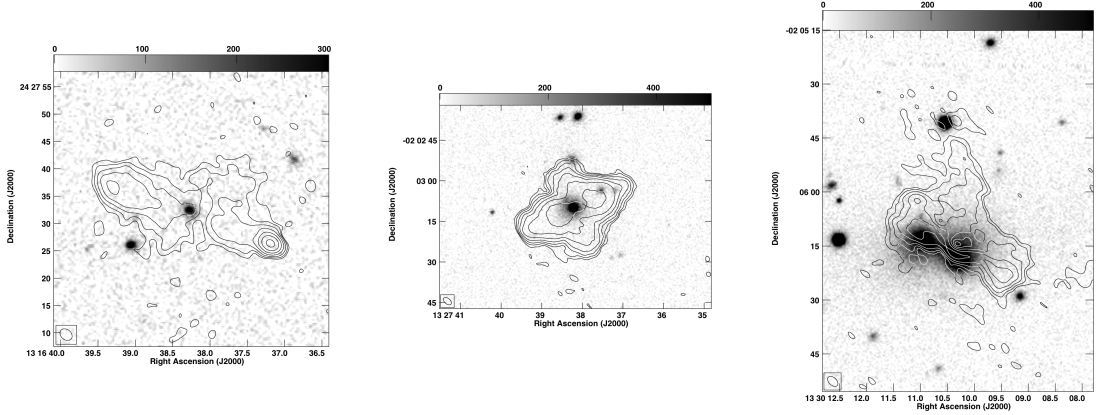

(a) J1316+2427          (b) J1327−0203          (c) J1330−0206

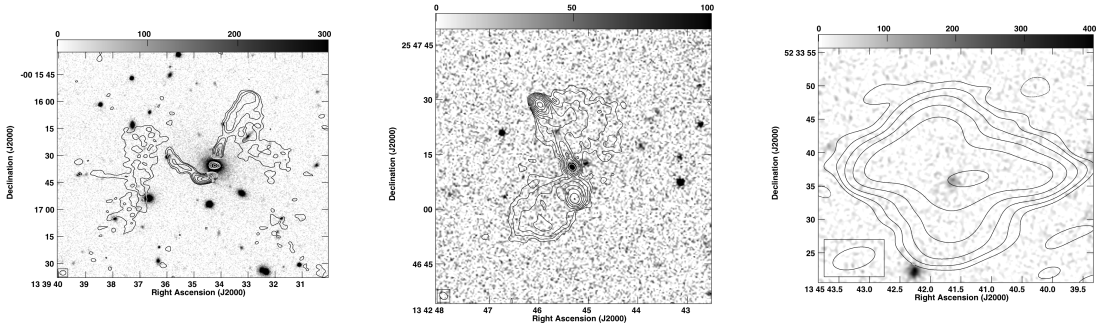

(d) J1339−0016          (e) J1342+2547          (f) J1345+5233

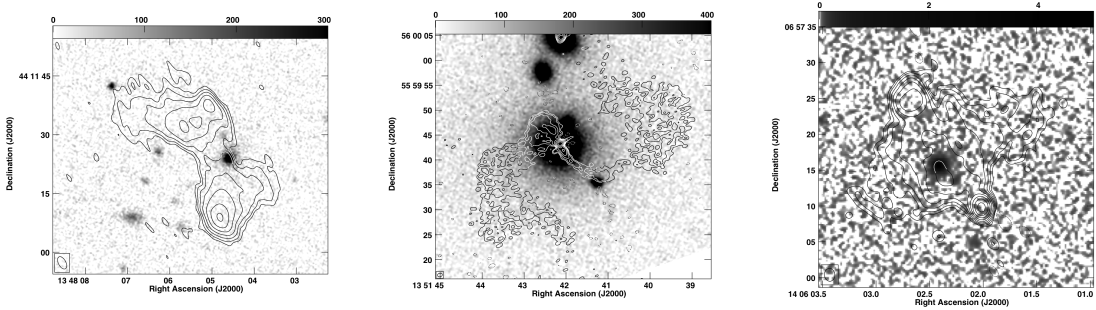

(g) J1348+4411          (h) J1351+5559          (i) J1406+0657

Fig. 35.—: Optical overlays of a sample of 100 low axial ratio radio sources (8/11).

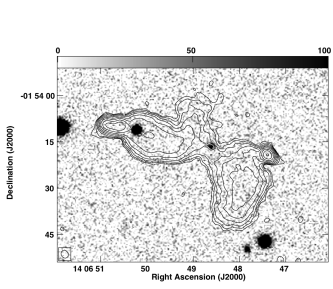

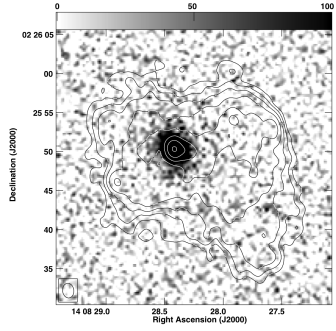

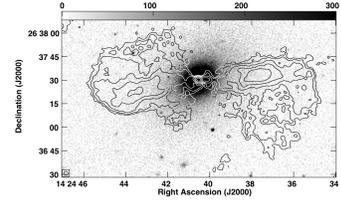

(a) J1406−0154

(b) J1408+0225

(c) J1424+2637

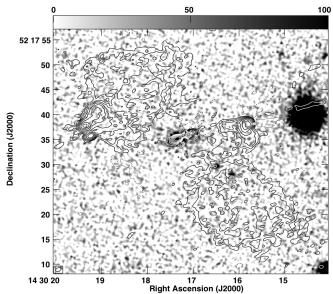

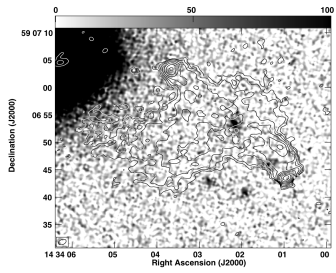

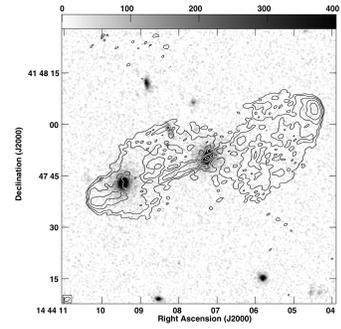

(d) J1430+5217

(e) J1434+5906

(f) J1444+4147

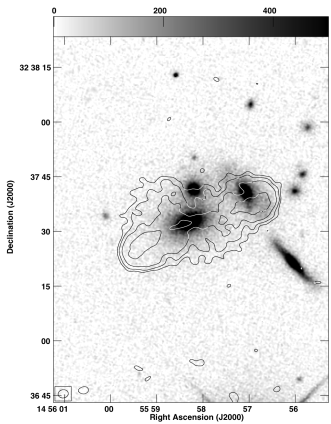

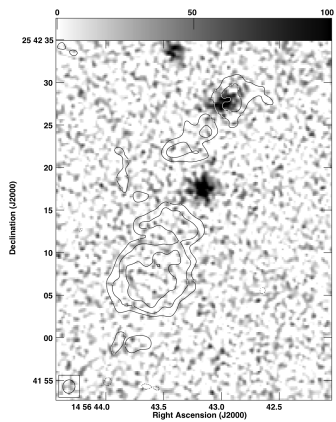

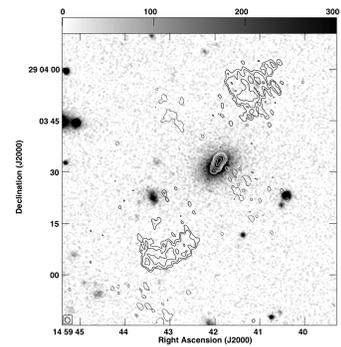

(g) J1455+3237

(h) J1456+2542

(i) J1459+2903

Fig. 36.—: Optical overlays of a sample of 100 low axial ratio radio sources (9/11).

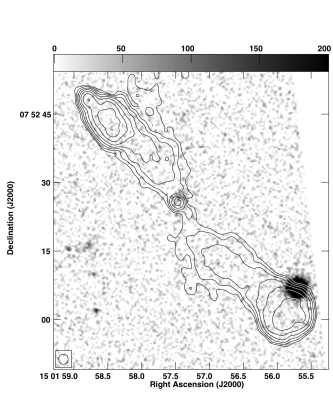

(a) J1501+0752

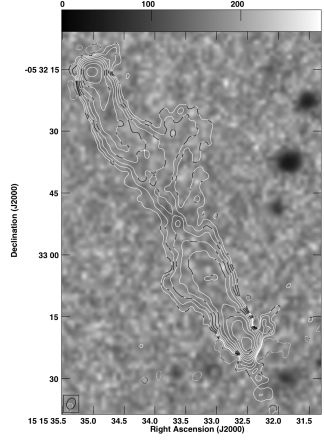

(b) J1515−0532

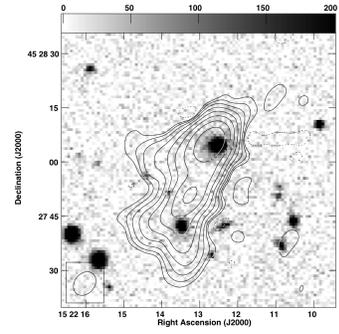

(c) J1522+4527

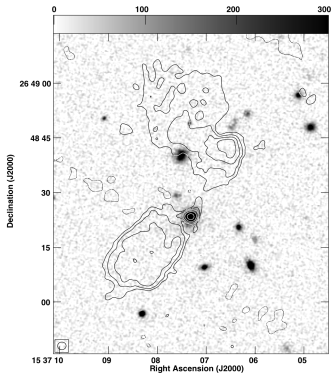

(d) J1537+2648

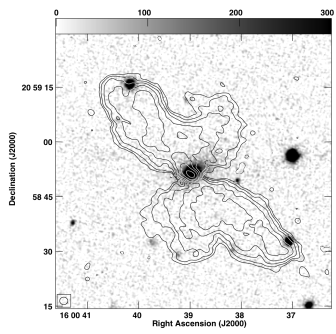

(e) J1600+2058

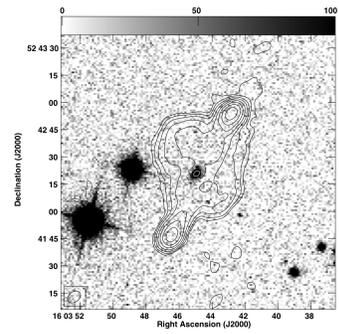

(f) J1603+5242

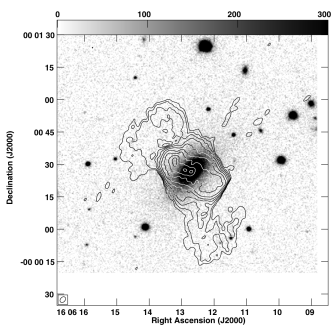

(g) J1606+0000

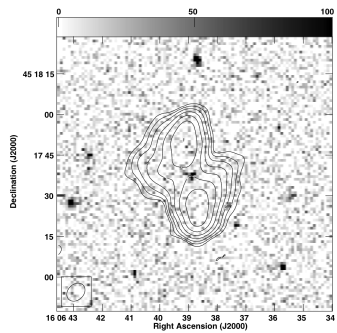

(h) J1606+4157

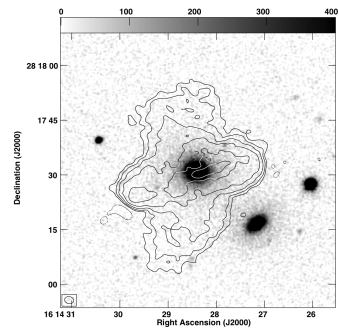

(i) J1614+2817

Fig. 37.—: Optical overlays of a sample of 100 low axial ratio radio sources (10/11).

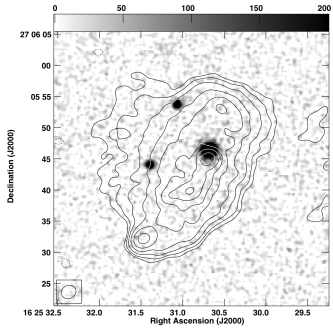 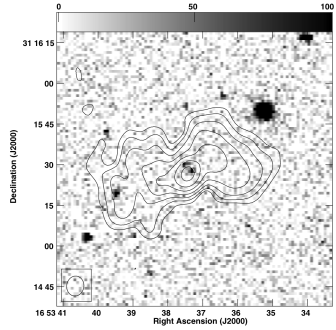 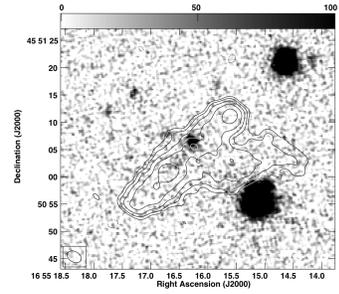

(a) J1625+2705     (b) J1653+3115     (c) J1655+4551

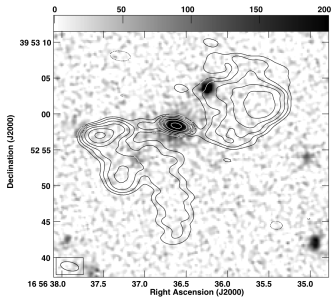 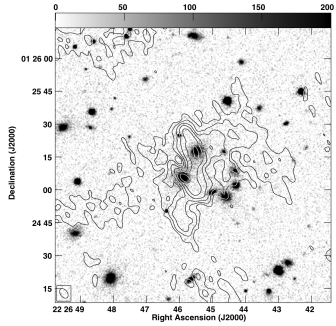 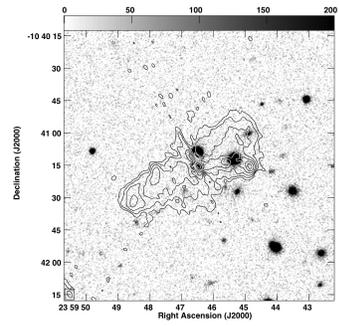

(d) J1656+3952     (e) J2226+0125     (f) J2359−1041

Fig. 38.—: Optical overlays of a sample of 100 low axial ratio radio sources (11/11).

Table 1.   Log of Observations

| Source | Archival[a] Band/Array | JVLA Band/Array, Date |
|--------|------------------------|------------------------|
| J0001−0033 | L/A | S/C, 07feb2016; S/B, 30may2016b; L/B, 03jun2016a |
| J0033−0149 | | S/C, 07feb2016; S/B, 30may2016; L/B, 03jun2016a |
| J0036+0048 | | S/C, 07feb2016; S/B, 30may2016; L/B, 03jun2016a |
| J0045+0021 | L/A;C/B | S/C, 07feb2016; S/B, 30may2016; L/B, 03jun2016a |
| J0049+0059 | L/A | S/C, 07feb2016; S/B, 30may2016; L/B, 03jun2016a |
| J0113+0106 | L/A; C/B | S/C, 07feb2016; S/B, 30may2016; L/B, 03jun2016a |
| J0115−0000 | | S/C, 07feb2016; S/B, 30may2016; L/B, 03jun2016a |
| J0143−0119 | L/A;C/B | S/C, 04feb2016c; S/B, 30may2016a; L/B, 31may2016 |
| J0144−0830 | L/A | S/C, 04feb2016c; S/B, 30may2016a; L/B, 31may2016 |
| J0145−0159 | L/A | S/C, 04feb2016c; S/B, 30may2016a; L/B, 31may2016 |
| J0147−0851 | | S/C, 04feb2016c; S/B, 30may2016a; L/B, 31may2016 |
| J0211−0920 | L/A | S/C, 04feb2016c; S/B, 30may2016a; L/B, 31may2016 |
| J0225−0738 | | S/C, 04feb2016c; S/B, 30may2016a |
| J0702+5002 | L/A | S/C, 04feb2016b; S/B, 28may2016a; L/B, 29may2016 |
| J0725+5835 | | S/C, 04feb2016b; S/B, 28may2016a; L/B, 29may2016 |
| J0805+4854 | | S/C, 04feb2016b; S/B, 28may2016a; L/B, 29may2016 |
| J0813+4347 | L/A | S/C, 04feb2016b; S/B, 28may2016a; L/B, 29may2016; L/A, 27dec2016 |
| J0821+2922 | L/A | S/C, 04feb2016a; S/B, 22may2016a; L/B, 28may2016b; L/A, 10jan2017 |
| J0836+3125 | | S/C, 04feb2016a; S/B, 22may2016a; L/B, 28may2016b |
| J0838+3253 | | S/C, 04feb2016a; S/B, 22may2016a; L/B, 28may2016b; L/A, 10jan2017 |
| J0845+4031 | L/A | S/C, 04feb2016a; S/B, 22may2016a; L/B, 28may2016b; L/A, 27dec2016 |

| Source | Archival[a] Band/Array | JVLA Band/Array, Date |
|---|---|---|
| J0846+3956 | L/A | S/C, 04feb2016a; S/B, 22may2016a; L/B, 28may2016b |
| J0859−0433 | L/A | S/B, 03jun2016b; L/B, 14aug2016 |
| J0914+1715 | | S/C, 04feb2016b; L/B, 30may2016c; S/B, 19jun2016; L/A, 10jan2017 |
| J0917+0523 | L/A | S/B, 03jun2016b; L/B, 14aug2016; L/A, 10jan2017 |
| J0924+4233 | L/A | S/C, 22apr2016; S/B, 22may2016a; L/B, 28may2016b |
| J0941−0143 | L/A | S/B, 03jun2016b; L/B, 14aug2016 |
| J0941+2147 | | S/C, 04feb2016b; L/B, 30may2016c; S/B, 19jun2016; L/A, 10jan2017 |
| J0943+2834 | | S/C, 04feb2016b; L/B, 30may2016c; S/B, 19jun2016; L/A, 10jan2017 |
| J1005+1154 | L/A | S/B, 03jun2016b; L/B, 14aug2016; L/A, 10jan2017 |
| J1008+0030 | L/A | S/B, 03jun2016b; L/B, 14aug2016; L/A, 10jan2017 |
| J1015+5944 | L/A | S/C, 22apr2016; S/B, 16jul2016 |
| J1040+5056 | | S/C, 22apr2016; S/B, 16jul2016 |
| J1043+3131 | L/A; C/B | S/C, 22apr2016; S/B, 16jul2016 |
| J1049+4422 | | S/C, 22apr2016; S/B, 16jul2016 |
| J1054+5521 | L/A | S/C, 22apr2016; S/B, 16jul2016 |
| J1055−0707 | | S/C, 29mar2016; S/B 24jul2016a; |
| J1102+0250 | | S/C, 29mar2016; S/B, 24jul2016a |
| J1111+4050 | L/A;C/B | . . . |
| J1114+2632 | | S/C, 22apr2016; S/B, 16jul2016 |
| J1120+4354 | | S/C, 22apr2016; S/B, 16jul2016; L/B, 26aug2016 |
| J1128+1919 | | S/B, 15jun2016; L/B, 26aug2016 |
| J1135−0737 | L/A | S/C, 29mar2016; S/B, 24jul2016a |

| Source | Archival[a] Band/Array | JVLA Band/Array, Date |
|---|---|---|
| J1140+1057 | | S/B, 15jun2016; L/B, 26aug2016 |
| J1200+6105 | | S/B, 22may2016a; S/B, 25jun2016 |
| J1201−0703 | | S/C, 29mar2016; S/B, 24jul2016a |
| J1202+4915 | L/A | S/B, 25jun2016 |
| J1206+3812 | L/A; C/B | S/B, 22jul2016 |
| J1207+3352 | L/A;C/B | S/B, 22jul2016 |
| J1210−0341 | L/A | S/C, 29mar2016; S/B, 24jul2016a |
| J1210+1121 | | S/B, 15jun2016 |
| J1211+4539 | L/A | . . . |
| J1218+1955 | | S/B, 22jul2016 |
| J1227−0742 | L/A | S/C, 29mar2016; S/B, 24jul2016a |
| J1227+2155 | L/A | S/C, 29mar2016; S/B, 22jul2016 |
| J1228+2642 | L/A | S/B, 22jul2016 |
| J1232−0717 | | S/C, 29mar2016; S/B, 24jul2016a |
| J1247+4646 | | S/B, 25jun2016 |
| J1253+3435 | L/A | S/B, 22jul2016 |
| J1258+3227 | | S/B, 22jul2016 |
| J1309-0012 | C/B | S/B, 24jul2016b |
| J1310+5458 | L/A | S/B, 25jun2016 |
| J1316+2427 | | S/B, 24jul2016b |
| J1327−0203 | L/A | S/B, 24jul2016b |
| J1330−0206 | | S/B, 24jul2016b |

| Source | Archival[a] Band/Array | JVLA Band/Array, Date |
|--------|------------------------|------------------------|
| J1339−0016 |  | S/B, 24jul2016b |
| J1342+2547 | L/A | S/B, 24jul2016b |
| J1345+5233 | L/A | S/C, 30mar2016a; S/AB, 25sep2016 |
| J1348+4411 | L/A | S/C, 30mar2016a; S/AB, 25sep2016 |
| J1351+5559 |  | S/C, 30mar2016a; S/AB, 25sep2016 |
| J1353+0724 |  | . . . |
| J1406−0154 | L/A | S/B, 24jul2016b |
| J1406+0657 | L/A | . . . |
| J1408+0225 | L/A | . . . |
| J1411+0907 |  | . . . |
| J1424+2637 |  | S/B, 24jul2016b |
| J1430+5217 | L/A | S/C, 30mar2016a; S/AB, 25sep2016 |
| J1433+0037 |  | . . . |
| J1434+5906 | L/A | S/C, 30mar2016a; S/AB, 25sep2016 |
| J1437+0834 |  | . . . |
| J1444+4147 |  | S/C, 30mar2016a |
| J1454+2732 |  | . . . |
| J1455+3237 |  | S/B, 24jul2016b |
| J1456+2542 | L/A | . . . |
| J1459+2903 | L/A; C/B | . . . |
| J1501+0752 |  | S/B, 05aug2016 |
| J1515+0532 |  | S/B, 05aug2016 |

| Source | Archival[a] Band/Array | JVLA Band/Array, Date |
|---|---|---|
| J1522+4527 | | S/C, 30mar2016a |
| J1537+2648 | | S/C, 30mar2016b; S/B, 05aug2016 |
| J1600+2058 | L/A | S/C, 30mar2016b; S/B, 05aug2016 |
| J1603+5242 | | S/C, 30mar2016b |
| J1606+0000 | L/A | S/B, 05aug2016 |
| J1606+4517 | L/A | S/C, 30mar2016b |
| J1614+2817 | L/A | S/B, 05aug2016 |
| J1625+2705 | L/A | S/B, 05aug2016 |
| J1653+3115 | | S/C, 30mar2016b |
| J1655+4551 | | S/C, 30mar2016b; S/B, 05aug2016 |
| J1656+3952 | L/A | S/C, 30mar2016b |
| J2226+0125 | | S/C, 04Feb2016; S/B, 30may2016a; L/B, 31may2016 |
| J2359−1041 | | S/C, 07feb2016; S/B, 30may2016; L/B, 03jun2016a |

[a]Roberts et al. 2015.

Table 2.   Figure Properties

| Source IAU Name | Figure Number | A-Array L-band[a] Low, Peak (mJy/beam) | B-Array C-band[a] Low, Peak (mJy/beam) | AB-Array S-band Low, Peak (mJy/beam) | B-Array L-band Low, Peak (mJy/beam) | B-Array S-band Low, Peak (mJy/beam) | C-Array S-band Low, Peak (mJy/beam) |
|---|---|---|---|---|---|---|---|
| J0001−0033 | 1 (a–d) | 0.10, 2.37 | ... | ... | 0.18, 6.82 | 0.09, 2.23 | 0.10, 10.2 |
| J0033−0149 | 1 (e–g) | ... | ... | ... | 0.18, 10.1 | 0.09, 3.72 | 0.09, 11.5 |
| J0036+0048 | 1 (h&i) | ... | ... | ... | 0.18, 65.8 | 0.09, 25.1 | ... |
| J0045+0021 | 2 (a–d) | 0.30, 178 | 0.50, 60.8 | ... | 0.18, 249 | 0.18, 109 | ... |
| J0049+0059 | 2 (e–h) | 0.09, 2.51 | ... | ... | 0.18, 9.60 | 0.09, 2.64 | 0.10, 12.9 |
| J0113+0106 | 2 (i), 3 (a–c) | 0.70, 18 | ... | ... | 0.18, 12.5 | 0.09, 4.32 | 0.14, 16.4 |
| J0115−0000 | 3 (d–f) | ... | ... | ... | 0.25, 15.2 | 0.09, 4.99 | 0.14, 15.6 |
| J0143−0119 | 3 (g–i), 4 (a&b) | 0.20, 44.4 | 0.09, 34.0 | ... | 0.28, 80.5 | 0.09, 47.0 | 0.16, 110 |
| J0144−0830 | 4 (c–f) | 0.09, 1.40 | ... | ... | 0.20, 7.25 | 0.09, 1.69 | 0.16, 5.88 |
| J0145−0159 | 4 (g–i), 5 (a) | 0.20, 2.56 | ... | ... | 0.25, 9.45 | 0.09, 3.02 | 0.20, 18.7 |
| J0147−0851 | 5 (b–d) | ... | ... | ... | 0.18, 33.2 | 0.09, 14.4 | 0.16, 30.2 |
| J0211−0920 | 5 (e–h) | 0.10, 2.81 | ... | ... | 0.18, 10.6 | 0.09, 2.65 | 0.13, 14.4 |
| J0225−0738 | 5 (i), 6 (a&b) | ... | ... | ... | 0.20, 51.4 | 0.09, 18.3 | 0.18, 49.1 |
| J0702+5002 | 6 (c–f) | 0.22, 5.74 | ... | ... | 0.18, 28.4 | 0.07, 4.88 | 0.09, 21.7 |
| J0725+5835 | 6 (g–i), 7 (a) | 0.22, 4.11 | ... | ... | 0.18, 10.0 | 0.07, 2.77 | 0.09, 8.19 |
| J0805+4854 | 7 (b–d) | ... | ... | ... | 0.18, 11.5 | 0.10, 4.38 | 0.06, 5.86 |
| J0813+4347 | 7 (e–h) | 0.14, 9.91 | ... | ... | 0.18, 22.1 | 0.07, 9.93 | 0.09, 17.1 |
| J0821+2922 | 7 (i), 8 (a–c) | 0.16, 1.93 | ... | ... | 0.17, 9.40 | 0.06, 1.98 | 0.09, 13.8 |
| J0836+3125 | 8 (d–g) | 0.16, 41.1 | ... | ... | 0.28, 60.0 | 0.09, 30.5 | 0.09, 53.5 |
| J0838+3253 | 8 (h&i), 9 (a&b) | 0.13, 5.52 | ... | ... | 0.20, 6.89 | 0.09, 6.54 | 0.09, 10.6 |
| J0845+4031 | 9 (c–f) | 0.20, 15.5 | ... | ... | 0.20, 31.7 | 0.09, 11.1 | 0.09, 32.3 |
| J0846+3956 | 9 (g–i), 10 (a) | 0.18, 16.1 | ... | ... | 0.20, 33.0 | 0.06, 11.1 | 0.09, 33.1 |
| J0859−0433 | 10 (b–d) | 0.20, 23.4 | ... | ... | 0.20, 26.8 | 0.07, 16.0 | ... |
| J0914+1715 | 10 (e–h) | 0.31, 157 | ... | ... | 0.35, 586 | 0.13, 160 | 0.18, 335 |
| J0917+0523 | 10 (i), 11 (a&b) | 0.28, 96.6 | ... | ... | 0.35, 165 | 0.14, 74.2 | ... |
| J0924+4233 | 11 (c–f) | 0.28, 6.24 | ... | ... | 0.17, 43.8 | 0.09, 4.74 | 0.10, 29.2 |
| J0941−0143 | 11 (g–i) | 0.60, 76.9 | ... | ... | 0.35, 188 | 0.14, 68.2 | ... |
| J0941+2147 | 12 (a–d) | 0.28, 54 | ... | ... | 0.25, 79.6 | 0.09, 39.5 | 0.13, 56.5 |
| J0943+2834 | 12 (e–h) | 0.20, 62.5 | ... | ... | 0.50, 142 | 0.13, 53.7 | 0.13, 91.3 |
| J1005+1154 | 12 (i), 13 (a&b) | 0.20, 2.72 | ... | ... | 0.28, 16.8 | 0.07, 3.57 | ... |
| J1008+0030 | 13 (c–e) | 0.20, 66.1 | ... | ... | 0.35, 75.2 | 0.08, 75.2 | ... |
| J1015+5944 | 13 (f–h) | 0.20, 98.1 | ... | ... | ... | 0.09, 49.3 | 0.10, 86.7 |

| Source IAU Name | Figure Number | A-Array L-band[a] Low, Peak (mJy/beam) | B-Array C-band[a] Low, Peak (mJy/beam) | AB-Array S-band Low, Peak (mJy/beam) | B-Array L-band Low, Peak (mJy/beam) | B-Array S-band Low, Peak (mJy/beam) | C-Array S-band Low, Peak (mJy/beam) |
|---|---|---|---|---|---|---|---|
| J1040+5056 | 13 (i), 14 (a) | ... | ... | ... | ... | 0.06, 2.20 | 0.07, 7.63 |
| J1043+3131 | 14 (b–e) | 0.40, 44.6 | 0.08, 36.2 | ... | ... | 0.09, 37.8 | 0.10, 68.0 |
| J1049+4422 | 14 (f&g) | ... | ... | ... | ... | 0.09, 4.50 | 0.10, 21.4 |
| J1054+5521 | 14 (h&i),15 (a) | 0.18, 14.1 | ... | ... | ... | 0.09, 9.40 | 0.10, 21.5 |
| J1055−0707 | 15 (b&c) | ... | ... | ... | ... | 0.09, 16.6 | 0.10, 42.9 |
| J1102+0250 | 15 (d&e) | ... | ... | ... | ... | 0.13, 83.0 | 0.10, 96.4 |
| J1111+4050 | 15 (f) | ... | 0.13, 11.3 | ... | ... | ... | ... |
| J1114+2632 | 15 (g&h) | ... | ... | ... | ... | 0.09, 26.8 | 0.07, 62.0 |
| J1120+4354 | 15 (i), 16 (a) | ... | ... | ... | ... | 0.09, 38.2 | 0.10, 71.0 |
| J1128+1919 | 16 (b&c) | ... | ... | ... | 0.25, 19.8 | 0.09, 15.1 | ... |
| J1135−0737 | 16 (d–f) | 0.20, 2.00 | ... | ... | ... | 0.09, 3.86 | 0.07, 7.39 |
| J1140+1057 | 16 (g&h) | ... | ... | ... | 0.25, 32.3 | 0.09, 14.0 | ... |
| J1200+6105 | 16 (i) | ... | ... | ... | ... | 0.05, 7.61 | ... |
| J1201−0703 | 17 (a&b) | ... | ... | ... | ... | 0.10, 8.29 | 0.10, 19.8 |
| J1202+4915 | 17 (c&d) | 0.15, 7.97 | ... | ... | ... | 0.05, 7.34 | ... |
| J1206+3812 | 17 (e–g) | 0.20, 47.0 | 0.15, 23.1 | ... | ... | 0.07, 24.7 | ... |
| J1207+3352 | 17 (h) | ... | ... | ... | ... | 0.07, 51.6 | ... |
| J1210−0341 | 17 (i), 18 (a&b) | 0.10, 2.95 | ... | ... | ... | 0.09, 5.28 | 0.07, 16.5 |
| J1210+1121 | 18 (c&d) | ... | ... | ... | 0.20, 17.3 | 0.09, 4.57 | ... |
| J1211+4539 | 18 (e) | 0.20, 43.6 | ... | ... | ... | ... | ... |
| J1218+1955 | 18 (f) | ... | ... | ... | ... | 0.14, 88.4 | ... |
| J1227−0742 | 18 (g–i) | 0.18, 0.86 | ... | ... | ... | 0.13, 3.12 | 0.09, 7.47 |
| J1227+2155 | 19 (a&b) | 0.18, 3.74 | ... | ... | ... | 0.10, 2.91 | ... |
| J1228+2642 | 19 (c&d) | 0.08, 4.35 | ... | ... | ... | 0.10, 5.66 | ... |
| J1232−0717 | 19 (e&f) | ... | ... | ... | ... | 0.10, 19.0 | 0.10, 53.4 |
| J1247+4646 | 19 (g) | ... | ... | ... | ... | 0.07, 3.23 | ... |
| J1253+3435 | 19 (h&i) | 0.09, 17.0 | ... | ... | ... | 0.14, 13.8 | ... |
| J1258+3227 | 20 (a) | ... | ... | ... | ... | 0.14, 120 | ... |
| J1309-0012 | 20 (b&c) | ... | 0.30, 113 | ... | ... | 0.20, 60.1 | ... |
| J1310+5458 | 20 (d&e) | 0.20, 11.3 | ... | ... | ... | 0.07, 7.72 | ... |
| J1316+2427 | 20 (f) | ... | ... | ... | ... | 0.10, 10.5 | ... |
| J1327−0203 | 20 (g&h) | 0.20, 20.9 | ... | ... | ... | 0.20, 23.6 | ... |
| J1330−0206 | 20 (i) | ... | ... | ... | ... | 0.10, 3.34 | ... |

Table 2—Continued

| Source IAU Name | Figure Number | A-Array L-band[a] Low, Peak (mJy/beam) | B-Array C-band[a] Low, Peak (mJy/beam) | AB-Array S-band Low, Peak (mJy/beam) | B-Array L-band Low, Peak (mJy/beam) | B-Array S-band Low, Peak (mJy/beam) | C-Array S-band Low, Peak (mJy/beam) |
|---|---|---|---|---|---|---|---|
| J1339−0016 | 21 (a) | ... | ... | ... | ... | 0.14, 6.48 | ... |
| J1342+2547 | 21 (b&c) | 0.30, 76.4 | ... | ... | ... | 0.09, 29.9 | ... |
| J1345+5233 | 21 (d–f) | 0.09, 0.92 | ... | 0.07, 2.47 | ... | ... | 0.18, 8.74 |
| J1348+4411 | 21 (g–i) | 0.07, 5.96 | ... | 0.06, 4.59 | ... | ... | 0.07, 24.0 |
| J1351+5559 | 22 (a&b) | ... | ... | 0.04, 4.61 | ... | ... | 0.07, 17.4 |
| J1406−0154 | 22 (c&d) | 0.30, 57.6 | ... | ... | ... | 0.09, 24.2 | ... |
| J1406+0657 | 22 (e) | 0.20, 71.4 | ... | ... | ... | ... | ... |
| J1408+0225 | 22 (f) | 0.10, 13.9 | ... | ... | ... | ... | ... |
| J1424+2637 | 22 (g) | ... | ... | ... | ... | 0.09, 9.05 | ... |
| J1430+5217 | 22 (h&i), 23 (a) | 0.20, 40.2 | ... | 0.08, 12.4 | ... | ... | 0.14, 80.8 |
| J1434+5906 | 23 (b–d) | 0.15, 47.3 | ... | 0.06, 23.4 | ... | ... | 0.14, 41.2 |
| J1444+4147 | 23 (e&f) | ... | ... | 0.10, 2.15 | ... | ... | 0.14, 17.9 |
| J1455+3237 | 23 (g) | ... | ... | ... | ... | 0.06, 1.16 | ... |
| J1456+2542 | 23 (h) | 0.11, 1.18 | ... | ... | ... | ... | ... |
| J1459+2903 | 23 (i), 24 (a) | 0.20, 8.67 | 0.10, 10.1 | ... | ... | ... | ... |
| J1501+0752 | 24 (b) | ... | ... | ... | ... | 0.07, 14.0 | ... |
| J1515+0532 | 24 (c) | ... | ... | ... | ... | 0.15, 70.3 | ... |
| J1522+4527 | 24 (d) | ... | ... | ... | ... | ... | 0.07, 17.3 |
| J1537+2648 | 24 (e&f) | ... | ... | ... | ... | 0.15, 3.57 | 0.10, 16.8 |
| J1600+2058 | 24 (g–i) | 0.20, 6.89 | ... | ... | ... | 0.10, 8.72 | 0.10, 21.3 |
| J1603+5242 | 25 (a) | ... | ... | ... | ... | ... | 0.20, 42.7 |
| J1606+0000 | 25 (b&c) | 0.20, 82.8 | ... | ... | ... | 0.25, 80.5 | ... |
| J1606+4517 | 25 (d&e) | 0.20, 3.22 | ... | ... | ... | ... | 0.10, 11.1 |
| J1614+2817 | 25 (f&g) | 0.20, 8.07 | ... | ... | ... | 0.14, 8.55 | ... |
| J1625+2705 | 25 (h&i) | 0.70, 2.80 | ... | ... | ... | 0.20, 41.8 | ... |
| J1653+3115 | 26 (a) | ... | ... | ... | ... | ... | 0.40, 74.1 |
| J1655+4551 | 26 (b&c) | ... | ... | ... | ... | 0.10, 5.30 | 0.14, 16.7 |
| J1656+3952 | 26 (d&e) | 0.20, 8.06 | ... | ... | ... | ... | 0.07, 20.8 |
| J2226+0125 | 26 (f–h) | ... | ... | ... | 0.28, 10.9 | 0.09, 4.12 | 0.20, 8.79 |
| J2359−1041 | 26 (i), 27 (a&b) | ... | ... | ... | 0.25, 12.3 | 0.13, 11.2 | 0.14, 18.8 |

[a]Roberts et al. 2015.